\documentclass[a4paper,10pt]{article}
\usepackage{amsmath,graphicx,subfigure}       
\renewcommand{\L}{\Lambda}
\newcommand\bsp{\begin{split}}
\newcommand\esp{\end{split}}
\newcommand\f{\phi}
\newcommand\g{\gamma}
\newcommand\gc{\gamma_5}
\renewcommand\d{\delta}
\newcommand\e{\epsilon}
\newcommand\m{\mu}
\newcommand\n{\nu}
\renewcommand\r{\rho}
\newcommand\s{\sigma}
\newcommand\p{\psi}
\newcommand\no{\nonumber}
\newcommand\oh{\frac{1}{2}}
\newcommand\pa{\partial}
\newcommand\ds[1]{\ooalign{$\hfil/\hfil$\crcr$#1$}}
\begin{document}
\begin{titlepage}
\begin{flushright}
IFUM 649/FT \\
December 1999 \\
\end{flushright}
\vspace{1.5cm}
\begin{center}
{\bf \large DIMENSIONAL RENORMALIZATION OF YUKAWA THEORIES }

{\bf \large VIA WILSONIAN METHODS }
\footnote{Work supported in part by M.U.R.S.T.}\\
\vspace{1 cm}
{ M. PERNICI} \\ 
\vspace{2mm}
{\em INFN, Sezione di Milano, Via Celoria 16, I-20133 Milano, Italy}\\
\vspace{0.6 cm}
{ M. RACITI and F. RIVA }\\
\vspace{2mm}
{\em Dipartimento di Fisica dell'Universit\`a di Milano, I-20133 Milano,
Italy}\\
{\em INFN, Sezione di Milano, Via Celoria 16, I-20133 Milano, Italy}\\
\vspace{2cm}
\bf{ABSTRACT}
\end{center}
\begin{quote}

 ~~ In the 't Hooft-Veltman dimensional regularization scheme it is
 necessary to introduce finite counterterms to satisfy chiral Ward
 identities. It is a non-trivial task to evaluate these counterterms
 even at two loops.

  We suggest the use of Wilsonian exact renormalization group
  techniques to reduce the computation of these counterterms to
simple master integrals.

We illustrate this method by a detailed study of a generic Yukawa
model with massless fermions at two loops.

\end{quote}
\end{titlepage}
\section*{Introduction}

The dimensional regularization scheme devised by 't Hooft and
Veltman  \cite{HV,tH} and later systematized by Breitenlohner and Maison 
\cite{BM} (BMHV) is the only 
dimensional regularization scheme which is known to be consistent in
presence of $\gc$. It gives the correct result for the axial
anomaly, at the price of breaking $d$-dimensional Lorentz symmetry
and chiral symmetry; while the former is easily recovered, it is a
non-trivial task to satisfy the chiral Ward identities.

More popular is the naive dimensional regularization scheme (NDR)
\cite{CFH}
which, although inconsistent, 
is much easier to use and can be handled safely in most practical situations.
For a review on the subject, see e.g. \cite{Bonn} .

Due to the relevance of higher loop computations in the
standard model, where it is difficult to guarantee the
consistency of the naive scheme, it is worthwhile to investigate
thoroughly  consistent renormalization schemes.

The difficulties encountered in computing systematically in the BMHV
sche\-me the
non-invariant counterterms  can be divided in three
categories:

i) it is a complicated (but generally solvable)
algebraic problem to satisfy all the Ward 
identities (modulo anomaly problems), 
determining these counterterms as  combinations
of Feynman integrals;

ii) it is an analytically difficult problem to evaluate these
counterterms,
which in general involve Feynman integrals with several masses and/or
momenta; explicit formulae with several masses at more than one loop
usually range from being quite complicated 
(see e.g. \cite{Remiddi}) to being as yet unknown; 

iii) it is burdensome to store these counterterms (which include
evanescent ones) and to compute the renormalization group beta
function using them \cite{Sch}.

By comparison, we recall that in a vectorial theory (or in NDR, 
whenever possible)
step (i) is trivial in the minimal subtraction (MS) scheme, 
since the Ward identities are automatically satisfied;  
step (ii) is much simpler, since the poles are much easier to compute
than the finite parts of the Green functions;
step (iii) is almost trivial, since in general multiplicative
renormalization holds; simple formulae for the beta and gamma 
renormalization group functions are available in the MS scheme \cite{tH}.

The short-cuts used to get the appropriate answer for the problem at
hand include:

I) in a few cases, like in the case of an  axial current insertion in
a vectorial gauge theory, it is sufficient to equate the axial vertex
to the corresponding vector vertex multiplied by $\g_5$ to be
sure that the chiral Ward identities are satisfied \cite{Trueman};

II) after solving the algebraic problem, avoid carrying out step (ii);
this is the philosophy of algebraic renormalization
( for a review see \cite{Sorella}), used generally
only for demonstrative purposes, but which can also be used for making
explicit computations, as shown in some two-loop examples
carried out recently \cite{Grassi}.

To our knowledge the only paper in which the counterterms are
determined systematically at one loop in the BMHV scheme
in a chiral gauge theory is \cite{Martin},
in which the Bonneau-Zimmermann \cite{Bonnid} identities are used.

There are several computations of one-loop counterterms 
in the BMHV scheme for particular processes in the
standard model \cite{Korner}, but no syste\-ma\-tic one-loop treatment has
been given in such a scheme.

In this paper we  describe a general method for simplifying step (ii)
and partly (iii),
i.e. we show that the counterterms can  be reduced to zero-momentum
Feynman integrals with the  same auxiliary mass in all propagators,
and that the beta function can be computed easily without making
a direct use of the counterterms.

The auxiliary mass technique in this form has been used in the past  
at  two or more loops \cite{vD,CMM,RVL} in conjunction  with
MS. Being the MS scheme mass-independent \cite{Collins}, the auxiliary mass technique
gives gauge-breaking terms that are polynomial
in the auxiliary mass and then 
can be easily treated;
however in chiral theories with BMHV the MS scheme cannot be used, and
it is not easy to disentangle the gauge-breaking terms.

We will show how to do this by  Wilsonian methods \cite{Wil}.

Wilsonian methods have been used by Polchinski \cite{Polchi}
to simplify the proof of renormalizability in $\phi^4$; this proof has
been further simplified and generalized to other cases 
\cite{Becchi,flow,PRR,PR};
in particular gauge theories have been renormalized
using the effective Ward identities introduced in \cite{Becchi}.
In \cite{PR} mass-independent renormalization has been 
studied with these Wilsonian methods; it has been also shown in this paper
that the Wilsonian effective action satisfies an effective
renormalization group equation, which is the analogous of the
effective Ward identities.

In this paper we propose the exploitation in the BMHV scheme of the 
effective Ward identities and effective renormalization group equation
to compute the finite counterterms and the beta function in terms
of Wilsonian Green functions at zero momenta and masses, which
are easy to evaluate.

As a simple testing ground for our proposal we renormalize
systematically at two loops the most general Yukawa theory with
massless Dirac fermions; so far only the simplest Yukawa model with
pseudoscalar coupling (without chiral symmetry) has been renormalized
at two loops in the minimal BMHV subtraction scheme
\cite{Sch}.

We impose Wilsonian mass-independent renormalization conditions
compatible with the rigid chiral Ward identities.
Our renormalization scheme is chosen to coincide with the minimal 
subtraction scheme
in the non-chiral Yukawa case (this choice is done to simplify step
(i), but is not essential in our method); 
in the general case it gives the same two-loop $\beta$ and
$\g$ functions as in the minimal naive scheme \cite{Mac}.
 
Finally we discuss the coupling with external gauge fields; we impose
Wilsonian renormalization conditions compatible with the effective
Ward identities at one loop. At two loop we check only the gauge two-point
function Ward identity and the Adler-Bardeen 
non-renormalization theorem \cite{AB}.

Our method is based on the use of a Wilsonian flow belonging to the
class
characterized by
the cut-off functions $K_{\L}^{(n)}(x) = (\frac{\L^2}{x + \L^2})^n$
with $n=2,3,\ldots$; these cut-off functions
separate the propagators into hard and soft parts.
For $n \geq 2$ the counterterms chosen at a particular $\L$
renormalize the flow for any $\L$.
 
We renormalize the theory at zero momenta and masses along the flow
$n=2$, using the $n=3$ case as a check on computations.

The counterterms are zero-momentum integrals with  
the same $\L$ in all the propagators, which
can be reduced to a single `master integral' at two loops
 using recursion relations (for a review  see \cite{CMM}).
We have used Mathematica \cite{Wolfram} to evaluate 
these integrals.

In the first section we show how the exact renormalization group
method can be used in connection with dimensional renormalization to
renormalize the Yukawa models.

In the second section, after reviewing the recursion formula for 
the massive zero-momentum two-loop integrals, 
we give the one and two loops results for the Yukawa models. 

In the third section we discuss the effective Ward identities for
the Yukawa model in presence of external gauge fields.

We end with a concluding section.

In the appendix we give the complete two-loop bare Yukawa action.

\section{Dimensional renormalization and the Wilsonian effective action}

\subsection{'t Hooft--Veltman regularization scheme}

We recall how gamma matrices are treated in the 't Hooft and Veltman 
\cite{HV} dimensional regularization scheme as elaborated by 
Breitenlohner and Maison \cite{BM} (BMHV).

We work in Euclidean space; $\d_{\mu \nu}$ is the Kronecker delta in
$d=4-\e$ dimensions; 

\begin{equation}
\d_{\mu \nu} p_{\nu} = p_{\mu}~~;~~\d_{\mu \mu} = d 
\end{equation}

The gamma matrices $\g_{\mu}$ satisfy the
relation
\begin{equation}
\{ \g_{\mu},\g_{\nu} \} = -2 \d_{\mu \nu} I
\end{equation}
where
\begin{equation}
tr I = 4 ~~;~~ I \g_{\mu} = \g_{\mu} I = \g_{\mu}
\end{equation}

In the BMHV scheme $O(d)$ invariance is broken and one introduces
$O(4)\times {O(d-4)}$ invariant tensors : the $(d-4)$--dimensional Kronecker delta
$\hat{\d}_{\mu \nu}$ and the $4$--dimensional antisymmetric tensor
$\e_{\nu \rho \sigma \tau}$ , satisfying
\begin{equation}
\hat{\d}_{\m \n}  \d_{\n \r} = \hat{\d}_{\m \r}~~;~~
\hat{\d}_{\m \m} = - \e  ~~;~~
\hat{\d}_{\m \n} \e_{\n \r \sigma \tau} = 0
\end{equation}
Moreover one defines the `evanescent' tensors
\begin{equation}
\hat{p}_{\m} \equiv \hat{\d}_{\m \n} p_{\n} ~~;~~
\hat{\g}_{\m} \equiv \hat{\d}_{\m \n} \g_{\n}
\end{equation}
The matrix $\g_5$ is defined by
\begin{equation}
\gc \equiv \frac{1}{4!} \e_{\m \n \r \s} \g_{\m}
\g_{\n} \g_{\r} \g_{\s} 
\end{equation}
satisfying
\begin{equation}
\{ \gc, \g_{\m} \} =
\{ \gc, \hat{\g}_{\m} \} = 2 \hat{\g}_{\m} \g_5 ~~;~~ \gc^2 = I
\end{equation}

In Euclidean space the reflection symmetry takes the place of
hermiticity. Reflection symmetry is an antilinear involution, 
under which 
\begin{equation}
\Theta \g_\m = - \g_\m ~~;~~\Theta \p (x) = \bar{\p}(x') \g_1 ~~;~~
\Theta \bar{\p}(x) = \g_1 \p(x')
\end{equation}
where $x^{'1}=-x^1$, $x^{'\m} = x^{\m}$ for $\m \neq 1$.

For a multiplet of fermionic fields, a general local marginal fermionic bilinear, involving a 
scalar ($A$), pseudoscalar ($B$), vector ($V_\m$) and
pseudovector ($A_\m$), all of them real fields, can be written as
\begin{equation}\label{dec}\bsp
&\int \bar{\p}(H_1 \g_{\m} +
H_2 [\g_{\m},\g_5] + H_3\hat{\g}_{\m}+
 i H_4 \hat{\g}_{\m} \gc ) \pa_{\m} \p + 
 \\
&\int \bar{\p} (i H_5 A +  H_6 \gc B + i H_7  \g_{\m} V_{\m} +
i H_8  [\g_{\m},\gc] A_{\m} +\\
& \qquad\;
i H_9 \hat{\g}_{\m} V_{\m}+H_{10}\hat{\g}_{\m} \gc A_{\m}) \p 
\esp\end{equation}
where $H_i$ are  
matrices over flavour indices. Reflection invariance requires that
$H_i$ are hermitian.

Green's functions in $d$ dimensions
are obtained 
performing the Dirac algebra in the Feynman graphs with the above
rules. Actually the analytic continuation to continuous dimensions is
defined
on the scalar coefficients of the Green's functions expanded on a
basis
for tensorial structures. For chiral theories, being $O(d)$ broken,
such a basis includes also `hatted' tensors.

The poles of the  Green's functions for $\e\to 0$ are removed
by local counterterms. Loop by loop the singular part of the counterterms   
must subtract exactly the pole part of the  Green's functions,
including
the 'hatted' components; the finite parts of the counterterms instead are not uniquely
determined by the renormalization conditions which constrain only their
$4$--dimensional part. 
In fact the $d \to 4$ prescription requires, besides to take the limit
$\e\to 0$, to set the hatted tensors to zero.

In order to fix completely the fermionic counterterms, in the decomposition
(\ref{dec}) we will choose that $H_3,\, H_4, \,H_9, \,H_{10}$ have vanishing
finite term in their Laurent expansion.
The same convention is assumed for the coefficients $b_{ij}$, $d^a_{ij}$ and
$d^{ab}_{ij}$
in the bilinear scalar counterterms,
which have the form
\begin{equation}\bsp\label{fermct}
&\oh\int \partial_\mu\phi_i \partial_\nu\phi_j (a_{ij} \d_{\mu\nu} +
b_{ij} \hat\d_{\mu\nu}) + V^a_\mu(\phi_i\partial_\mu\phi_j c^a_{ij}
+\phi_i\hat\partial_\mu\phi_j d^a_{ij})+ \\
&\qquad V^a_\mu V^b_\mu \phi_i
c^{ab}_{ij}\phi_j+V^a_\mu \hat\d_{\mu\nu} V^b_\nu \phi_i  d^{ab}_{ij}\phi_j
\esp
\end{equation}

\subsection{Yukawa models }

Consider the most general Yukawa model with massless fermions.

In dimensional regularization the bare action is 
\begin{equation}\label{bareY}
S = \int \frac{1}{2} \phi_i c_{ij}\phi_j + \bar{\psi}c\psi + 
\mu^{\e /2} \bar{\psi} c_i\psi \phi_i +  
 \frac{\mu^{\e }}{4!} c_{ijkl}\phi_i \phi_j \phi_k\phi_l
\end{equation}
and it is chosen to be reflection symmetric.

At tree level
\begin{eqnarray}\label{cc0}
&&c_{ij}^{(0)}(p) = \d_{ij} p^2 + m^2_{ij}~~;~~
c^{(0)}(p) = i \g_{\mu}p_{\mu} \nonumber \\
&&c_{i}^{(0)} = i y_i ~~;~~ c_{ijkl}^{(0)}= h_{ijkl}
\end{eqnarray}
We define
\begin{eqnarray}\label{ySP}
y_i = S_i I+ i P_i \gc ~~;~~ Y_i = S_i + i P_i
\end{eqnarray}
The constants $c,c_i$ are matrix-valued, $c =
c_{IJ}$ , etc., where ${\scriptstyle I,J}$ are the internal indices of the fermions
$\psi_I$.  The matrices $S_i,P_i$  are hermitian. 

We will consider a group $G$ which is not necessarily semi-simple,
with structure constants $f^{abc}$.
The fields transform under   linear (chiral)
representations of the group which are in  general reducible:
\begin{equation}\label{group}
\psi \to g \psi ~~; ~~ \bar{\psi} \to \bar{\psi} \bar{g}^{-1} ~~; ~~ 
\phi_i \to h_{ij} \f_j 
\end{equation}
where 
\begin{equation}\label{group2}
g = \exp(i \e^{a} t^{a}) ~~; ~~\bar{g} = \exp(i \e^{a} \bar{t}^{a})~~; ~~h = \exp(i \e^{a} \theta^{a})
\end{equation}
and
\begin{equation}\bsp
&t^a = t_R^a P_R + t_L^a P_L = t_s^a + t_p^a \gc
~~;~~ 
\bar{t}^a = t_L^a P_R + t_R^a P_L =
 t_s^a - t_p^a \gc  \\
&
P_R = \oh (1+\gc) ~~;~~
P_L = \oh (1-\gc)
\esp\end{equation} 
$t_R^a$ and $t_L^a$ belong in general to different representations of
$G$. The scalar fields are in a real representation; then $\theta^a$
are antisymmetric imaginary matrices.

The Yukawa coupling $\bar{\p} y_i \p \f_i$\,
 is invariant under these transformations provided
the following relations hold 
\begin{equation}\label{yinv}
y_j \theta^a_{ji} + y_i t^a - \bar{t}^a y_i = 0 ~~;~~
y_j^\dagger \theta^a_{ji} + y_i^\dagger \bar{t}^a - 
t^a y_i^\dagger = 0 
\end{equation}
or equivalently
\begin{equation}
Y_j \theta^a_{ji} + Y_i t_R^a - t_L^a Y_i = 0 ~~;~~
Y_j^\dagger \theta^a_{ji} + Y_i^\dagger t_L^a - t_R^a Y_i^\dagger = 0
\end{equation}

The tree-level action  is invariant under these
chiral transformations, apart from an evanescent fermionic kinetic
term. Higher corrections in the bare action will require also non-invariant
counterterms; in the next subsection we will discuss how Wilsonian
methods can be applied to determine the counterterms in order to preserve the Ward identities.

\subsection{Wilsonian effective action}\label{subsec:wilseff}

The bare action (\ref{bareY}) has the form 
\begin{eqnarray}\label{bare}
S (\Phi)= \frac{1}{2}\Phi D^{-1}\Phi + S^I(\Phi) 
\end{eqnarray}
where we use a compact notation in which $\Phi$ is a collection of fields.
$S^I$ contains the tree-level interaction terms and the counterterms.

Consider an analytic cut-off function 
\begin{eqnarray}\label{Kn}
K_{\L}(p,M) = ( \frac{\L^2}{p^2+M^2+\L^2} )^n
\end{eqnarray} 
for $n \geq 2$ ; $M^2 = m^2 \geq 0$ in the scalar sector; $M^2=0$ in the
fermionic sector.

Let us split the propagator in two parts characterized by a scale 
$\L > 0$; defining
\begin{eqnarray}
D_H = D_{\L} = D(1-K_{\L}) ~~;~~ D_S = D K_{\L}
\end{eqnarray}

Let us make an `incomplete integration'  over the hard modes 
\begin{equation}\label{Zj}
Z_{\L}[J] = exp \frac{1}{\hbar} W_{\L}[J] \equiv 
 \int{\cal D}\Phi 
e^{-\frac{1}{ \hbar}[S_{\L} (\Phi) - J \Phi]}  
\end{equation}
with bare action 
\begin{eqnarray}\label{lbare}
S_{\L}(\Phi)= \frac{1}{2} \Phi D_{\L}^{-1}\Phi + S^I(\Phi) 
\end{eqnarray}
The Green functions obtained from 
$Z_{\L}[J]$ are infrared finite for $\L > 0$ even at exceptional momenta.

The flow of this functional from $\L$ to zero can be represented as 
\begin{equation}\label{ZL}
Z[J] = Z_{0}[J] =  
e^{\frac{\hbar}{2} \frac{\d}{\d J} D_{\L}^{-1} K_{\L}
\frac{\d}{\d J}} Z_{\L}[J]
\end{equation}

The Wilsonian theory at scale $\L$ has a bare action differing from
the one of the usual theory at $\L =0$ by the term 
\begin{equation}\label{Delta}
S_{\L}(\Phi) - S(\Phi) = \frac{1}{2}\Phi D_{\L}^{-1}K_{\L} \Phi
\end{equation}
which has ultraviolet dimension less or equal to zero, due to the above 
made choice of the cut-off function $K_{\L}$; the renormalization of the
usual theory implies the renormalization of the Wilsonian theory and
viceversa.

To impose the  renormalization conditions it is useful to separate
the Wilsonian $1$-PI functional generator,
obtained by making a Legendre transformation on $W_{\L}[J]$,
 into the quadratic tree-level and interacting parts:
\begin{eqnarray}
\Gamma_{\L}[\Phi] = \frac{1}{2} \Phi D_{\L}^{-1} \Phi +
\Gamma_{\L}^I[\Phi]
\end{eqnarray}
and to make a Taylor expansion of $\Gamma_{\L}^I[\Phi]$ in fields,
$d$-dimensional momenta and masses; 
the four-dimensional terms in this expansion form a local  functional,
which will be called $S_W$.

Let us now discuss briefly why the renormalization program can be
performed on the $Z_{\L}$ functional with some advantages:

i) since the hard propagator $D_H(p)$ is regular at $p=0$  there are
no infrared problems in the evaluation of the counterterms, so
that the counterterms can be chosen to be Feynman integrals at zero
momenta and masses using  a mass-independent Wilsonian renormalization
scheme \cite{PR};

ii) for a Wilsonian flow with $n \geq 2$ it is simple to prove that
the renormalization at a fixed $\L$ implies the  finiteness 
of the Wilsonian effective action at every value $\L$ of the Wilsonian
flow, in particular at $\L = 0$; moreover because the  propagators $D_\L$
at $\L = 0$ do not depend on $n$, a bare action that renormalizes a
given flow will also make finite the flows for every $n\geq 2$.  

iii) for a Wilsonian flow with $n \geq 2$
the Ward  identities and the renormalization group equation on
the functional $Z_{\L=0}$ are equivalent to the corresponding
effective Ward  identities \cite{Becchi,PRR}  and 
effective renormalization group equation \cite{PR} on
$Z_{\L}$, so that choosing renormalization conditions compatible
with the effective Ward identities the validity of the
Ward identities on $Z_{\L=0}$ follows; furthermore 
the renormalization group beta function can be expressed in terms
of Feynman integrals at zero momenta and masses.

These considerations hold for any renormalizable theory in four
dimensions; we consider here only the Yukawa model, which is the
simplest chiral theory. 
Actually its rigid Ward identities are too  simple to illustrate
point (iii), since $Z_{\L}$ satisfies the same 
rigid Ward identities;
however introducing external currents one has to study the local
Ward identities, whose effective form will be used  in a later section.

In the Yukawa model the  bare constants in the action are chosen of the form 
\begin{equation}\label{ct}\bsp
&c_A =  \sum_{l \geq 1} \hbar^l N_d^l
c^{(l)}_A(\e) \\
& N_d = (4 \pi)^{\e/2-2} \Gamma (1+\frac{\e}{2})~~;~~
c^{(l)}_A(\e) = \sum_{r \geq 0} c^{(l)}_{A,-r} \e^{-r}
\esp\end{equation} 

We use a modified subtraction scheme \cite{BL} , 
in which $N_d$ is introduced in
order to avoid $\g_E~,~ ln(4\pi)$ and $\pi^2$ factors. 
We choose a mass-independent renormalization scheme, i.e. all the bare
terms, apart from $c_{ij}$, are independent from $m^2$; the latter term
depends polynomially on it.

The bare constants must be fixed by suitable renormalization
conditions on the Wilsonian effective action.

In the Yukawa model the marginal part 
\footnote{In a Wilsonian
 mass-independent scheme, as discussed in \cite{PR}, the
 renormalization of the mass parameter is treated in a way very
 similar to the
kinetic term and therefore is included in the marginal part of the
 action.} 
$S_W^\L$
 of the Wilsonian
effective action is defined according to:
 
\begin{equation}\label{last}\bsp
& \Gamma^\L = S_W^\L+\Gamma^\L_{irr}\\ 
&S_W^\L = \int
\bar{\psi} \g_\m a \pa_\m \psi -
\oh a_{ij} \phi_i \pa^2 \phi_j + \oh \phi_i m^2_{rs}a^{rs}_{ij}\phi_j +\\
&\qquad\qquad\bar{\psi} a_i \psi \phi_i +
\frac{1}{4!} a_{ijkl} \phi_i \phi_j \phi_k \phi_l
\esp\end{equation}
$\Gamma_{irr}$ contains all the terms of dimension greater than four in the
expansion of $\Gamma_\L$
in the fields, momenta and masses.

The renormalization conditions fix the limit for $\e\to 0$ of the functions  $g_A = \{a, a_{ij}, a^{rs}_{ij}, a_i, a_{ijkl}\}\,
$ at $\L = \m $  and perturbatively determine uniquely the bare action, as discussed in the first section.

From the Feynman graphs rules the terms $g_A$ at order $l$ in the
loop expansion are
\begin{equation}\label{marg} 
g_A^{(l)}= \sum_{r=0}^{l} (\frac{\mu }{ \L})^{r \e} g^{(r,l-r)}_A
\end{equation}
in which the dependence on $\Lambda $ and $\mu$ is made explicit;
 $g^{(l,0)}_A$ is the $l$-loop graph contribution,
while $g^{(r,l-r)}_A$ , $r=0,...,l-1$
is the contribution due to the $r$-loop graphs with 
counterterms of overall loop order $l-r$; 
$g^{(r,l-r)}_A$ are independent from $\mu$ and $\L$ . 
 
Being the theory renormalized, one gets, for $\e  \to 0$ 
\begin{equation}\label{res}
\mu \frac{\partial}{\partial \mu } g^{(l)}_A  =
-\L \frac{\partial}{\partial \L } g^{(l)}_A  \to
\sum_{r=1}^{l} r~ g^{(r,l-r)}_{A,-1}
\end{equation}

Taking $s$ derivatives with respect to
$\L$ in (\ref{marg}) one has still a finite expression for 
$\e \to 0$, so that one obtains the consistency conditions 
\begin{equation}\label{cons}
\sum_{r=1}^{l} r^s~ g^{(r,l-r)}_{A,-n} = 0
\end{equation}
for $n=2,...,l$ and $s=1,...,n-1$. 

For the two-loop case the consistency condition (\ref{cons}) for $n=2$
and $s=1$ is known to hold
for each Feynman graph and its counterdiagrams \cite{FO},
providing a useful check.

To renormalize the theory we assign finite values to the constants
$g_A$ at $\L = \mu$
\begin{eqnarray}\label{grf}
g_A = r_A + O(\e)
\end{eqnarray}

Each coefficient in the Laurent expansion in $\e$ must be determined
to establish
completely the renormalization scheme. In a subtraction scheme in
which the bare constants are fixed to have all positive powers of $\e$
equal to zero (see eq. (\ref{ct})) and the renormalized quantities are
fixed to be non singular and with determined $\e^0$ coefficient, the
renormalization scheme is completely fixed. The $O(\e)$ term in
eq. (\ref{grf})
refers to the fact that one cannot fix the positive powers of $\e$
both
in the renormalization condition and in the bare action.

In order to respect the rigid Ward identities one must choose
for $r_A$ 
group-covariant quantities under the chiral transformations. They can
be constructed out of the tree coupling constants $y$ and $h_{ijkl}$,
taking into account the invariance properties 
\begin{equation}\label{chir}
\bar{g}^{-1} y_i g = h_{ij}y_j~~;~~ 
g^{-1} y_i^{\dagger} \bar{g}=h_{ij}y^\dagger_j
\end{equation}
Furthermore one can examine which choice for $r_A$ gives the simplest
expression for the counterterms.

For the non-chiral theory, in which the couplings $P_i$ of
 eq.(\ref{ySP})
 are vanishing
and $g = \bar g = g_s $, the simplest counterterms are those
determined by 
the MS scheme, corresponding to renormalization conditions
$r_A(S)$
that can be explicitly computed. 

For the chiral theory  we will define the counterterms  as suitable 
functions $c_A(S,P)$ which, for $P=0$,
coincide with the corresponding functions of the non chiral case.
To this aim loop by loop we will choose $r_A $ applying a covariantization formula
to  the corresponding $r_A(S)$. 
A comparison of eq.(\ref{chir}) with the analogous equation in the non chiral
case with the same group $G$:
\begin{equation}\label{nonchir}
g_s^{-1} S_i\,g_s=h_{ij}S_j
\end{equation}
suggests the recipe 
for the Green's functions of interest : 
\begin{equation}\bsp
&S_{i_1}S_{i_2}\ldots S_{i_{2n+1}}I\to 
y_{i_1}y^\dagger_{i_2}\ldots y_{i_{2n+1}}\\
&\g_\mu S_{i_1}S_{i_2}\ldots S_{i_{2n}}\to
\g_\mu y^\dagger_{i_1}y_{i_2}\ldots y_{i_{2n}} \\
&Tr\,\left[S_{i_1}S_{i_2}\ldots S_{i_{2n}}\right]\to 
\frac{1}{4}Tr\,tr\,\left[y_{i_1}y^\dagger_{i_2}\ldots y^\dagger_{i_{2n}}\right]=
\\&\qquad\qquad\qquad\qquad\quad
\frac{1}{2} Tr\,\left(
Y_{i_1} Y^{\dagger}_{i_2} ... Y_{i_{2n-1}} Y^{\dagger}_{i_{2n}} +
Y^{\dagger}_{i_1} Y_{i_2} ... Y^{\dagger}_{i_{2n-1}} Y_{i_{2n}} \right) 
\esp\end{equation}
where $Tr$ denotes the trace over the internal fermionic indices.

As an example let us consider the renormalization condition to be
imposed on the fermionic self--energy.
In the non chiral theory, using the minimal subtraction one computes:
\begin{equation}
\Sigma(p)|_{marg} = i \ds{p} \;r(S)
\end{equation}
Using the covariantization procedure $i \ds{p} \;r(S)$ must be
replaced by $ i \ds{p}\; r(y,y^\dagger)$.
In order to be compared with the standard form of eq.(\ref{dec}) $\Sigma$ can be written as
\begin{equation}\label{sigma}
\bsp
\Sigma(p)|_{marg} =\frac{i}{2}\, \Big\{&\ds{p}
\left[r(Y,Y^\dagger)+r(Y^\dagger,Y)\right]+\\&\oh\left[\ds{p},\gc\right]\left[r(Y,Y^\dagger)-r(Y^\dagger,Y)\right]\Big\}
\esp
\end{equation}
which differs from $ i \ds{p}\; r(y,y^\dagger)$ by evanescent terms.
Eq.(\ref{sigma}) gives the renormalization condition for the
chiral theory. 
Observe that, if the theory is reflection symmetric, $r(y,y^\dagger)$
is hermitian.

The counterterms $c_A$ are completely determined by the pole part and by
the constant part of the corresponding vertices. They can be
decomposed into two parts 
\begin{equation}\label{ccc}
c_A = c_A^{NDR} + \Delta c_A
\end{equation}
where $c_A^{NDR}$ has the same group structure as $r_A$ and, due to
the choice of renormalization conditions,
has only pole part in $\e$ (see comment after eq.(\ref{grf})).

The remaining part $\Delta c_A$ vanishes in the non-chiral case
($P_i = 0$).

The non-marginal relevant terms satisfy  
\begin{eqnarray}\label{rc2}
\Gamma_{ij}^{\L = 0}|_{p = m=0} = 0 ~~;~~
\Gamma_{IJ}^{\L = 0}|_{p = m=0} = 0 
\end{eqnarray}

In dimensional regularization, being the massless tadpoles equal to
zero, eq.(\ref{rc2}) corresponds to have
vanishing
bare relevant counterterms.
\section{Explicit computations}

\subsection{Master integrals}

Using the cut-off function (\ref{Kn}) the hard propagator has the form
\begin{equation}
D_{\L}(p,M) = 
D(p,M) \frac{p^2 + M^2}{ \L^2} \sum_{r=1}^n \left( \frac{\L^2}{ (p^2+M^2+\L^2} \right)^r
\end{equation}
The counterterms are computed in terms of Wilsonian Green
functions at zero momenta and masses $M = 0$; 
the corresponding  Wilsonian
Feynman graphs with $I$ internal lines have $I$ sums 
$\sum_{r_1=1}^n ...\sum_{r_I=1}^n$.

Since $D(p,M) (p^2+M^2)$ is polynomial in $p$  the counterterms 
are expressed in terms of massive zero momentum tensor integrals, in which all
the propagators have `mass' $\L$.
Taking traces one can reduce these tensor
integrals in terms of scalar integrals. 

At one loop the scalar integrals are
\begin{equation}
I_a = \int \frac{d^dq_1 }{ N_d \pi^{d/2}} \frac{1}{ (q_1^2 + 1)^a}
\end{equation}
At two loops  the scalar integrals have the form
\begin{equation}
I_{a,b,c} = \int \frac{d^dq_1}{ N_d \pi^{d/2}} \int 
\frac{d^dq_2}{ N_d \pi^{d/2}}
 \frac{1}{ (q_1^2 + 1)^a (q_2^2 + 1)^b ((q_1+q_2)^2 + 1)^c }
\end{equation}
and so on at higher loops.

At one loop one has the well-known explicit expression
\begin{equation}
I_{a} = \frac{ \Gamma (a-d/2)}{ N_d ~\Gamma (a) }
\end{equation}

At higher loops an exact expression for these integrals is not known; 
they can be reduced, using recursion relations, to a small number of
master integrals which can be expanded in $\e$; to renormalize at 
$l$-loop one must know the $I^{(l)}$ up to the $O(1)$ term, 
the $I^{(l-1)}$ up to the $O(\e)$ term 
and so on (in minimal subtraction it is sufficient to
know the $I^{(l)}$ up to the $O(1/\e)$ term,
the $I^{(l-1)}$ up to the $O(1)$ term and so on; however in presence of
chiral symmetries MS is not sufficient).

At two loops there is only one master integral (for a review see e.g.
\cite{CMM}); 
it is known how to compute the finite parts of the three-loop master integral
 \cite{RVL}. 

In this paper we will restrict to two-loop computations; 
the recursion relation is obtained from the identity
\begin{equation}
\int \frac{d^dq_1}{ N_d \pi^{d/2}} \int 
\frac{d^dq_2}{ N_d \pi^{d/2}}
\frac{\partial}{ \partial q_1^{\mu}}  
 \frac{q_1^{\mu}}{ (q_1^2 + 1)^a (q_2^2 + 1)^b ((q_1+q_2)^2 + 1)^c } = 0
\end{equation}
which implies
\begin{equation}\bsp
&(d-2b-c)I_{a+1,b,c}- c(I_{a,b-1,c+1}-I_{a-1,b,c+1}) +\\&
2b I_{a,b+1,c} + c I_{a,b,c+1} = 0
\esp\end{equation}
From this relation and the fact that 
$I_{a,b,c}$ is totally symmetric in its indices it follows the
recursion relation
\begin{equation}\label{recur}\bsp
3 \,a\, I_{a+1,b,c} =\, &c\, (I_{a-1,b,c+1} - I_{a,b-1,c+1} ) +\\&
b \,( I_{a-1,b+1,c} - I_{a,b+1,c-1}) + (3 a - d) I_{a,b,c}
\esp\end{equation}
Using this recursion relation
one can express all $I_{a,b,c}$ , for $a,b,c > 0$ 
in terms of the master integral
$I_{1,1,1}$ and in terms of $I_{a',b',c'}$, with one of the indices vanishing;
in the latter case the two-loop integral reduces to the product of two 
one-loop integrals:
\begin{equation}
I_{a,b,0} = I_a I_b
\end{equation}
One has
\begin{equation}\bsp
& I_{1,1,1} = - \frac{6}{ \e^2} - \frac {9}{ \e} + 
\frac{3}{ 2} \left(v-5\right) + O(\e)  \\
& v \equiv 3 \lim_{\e \to 0} I_{2,2,2} = - 2
-\frac{4}{\sqrt{3}} \int_0^{\pi/3}dx ~ln \left(2 sin \frac{x}{2}\right) \simeq 0.34391
\esp\end{equation}
These recursion relations can be solved very fast by computer in
the cases of interest.
Recently all the Laurent series of $I_{1,1,1}$ has been computed in
\cite{Davyd}.
\subsection{One-loop results}

At one loop the renormalization conditions (\ref{grf})
on the Wilsonian vertices are  
\begin{equation}\label{rc1}\bsp
&r_{ij}^{(1)} = r^{\phi}_{(1)} Y_{ij} \qquad\;\,;\qquad r_{ij}^{(1)rs}= 
r^{m^2}_{(1)} h_{ijrs} \\
&r^{(1)} =  r^{\psi}_{(1)} y_i^{\dagger}  y_i \qquad;\qquad
r_i^{(1)} = i r^{v3}_{(1)} y_j y_i^{\dagger} y_j \\ 
&r_{ijkl}^{(1)} = r^{v4}_{(1)1} Y_{(ijkl)}
+ r^{v4}_{(1)2} h_{(ij}^{rs} h^{rs}_{kl)} \esp
\end{equation}
where the symmetrizations $(\ldots)$ are with weight one: e.g.
$h_{(ijkl)} = h_{ijkl}$; we defined 
\begin{equation}\label{defY}
Y_{i_1i_2...i_{2n-1}i_{2n}} \equiv \frac{1}{2} Tr\,\left(
Y_{i_1} Y^{\dagger}_{i_2} ... Y_{i_{2n-1}} Y^{\dagger}_{i_{2n}} +
Y^{\dagger}_{i_1} Y_{i_2} ... Y^{\dagger}_{i_{2n-1}} Y_{i_{2n}} \right)
\end{equation}

The same parametrization holds for the bare constants $c_A^{NDR}$
defined in eq.(\ref{ccc}).

\begin{figure} 
\centering
	\subfigure[]{
		\label{fig:subfig:a}
        	\includegraphics[width=2cm]{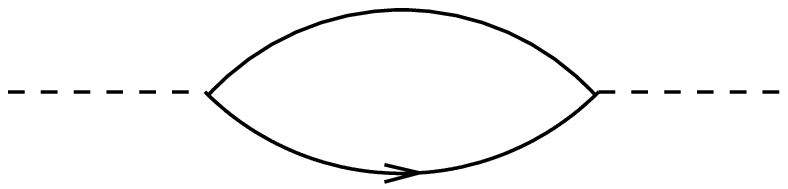}}
	\hspace{0.4cm}
	\subfigure[]{
 		\label{fig:subfig:b}
        	\includegraphics[width=2cm]{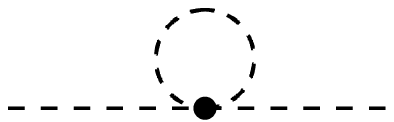}}
	\hspace{0.4cm}
	\subfigure[]{
		\label{fig:subfig:c}
		\includegraphics[width=2cm]{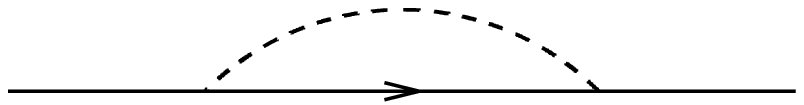}}
	\hspace{0.4cm}
	\subfigure[]{
		\label{fig:subfig:d}
		\includegraphics[width=2cm]{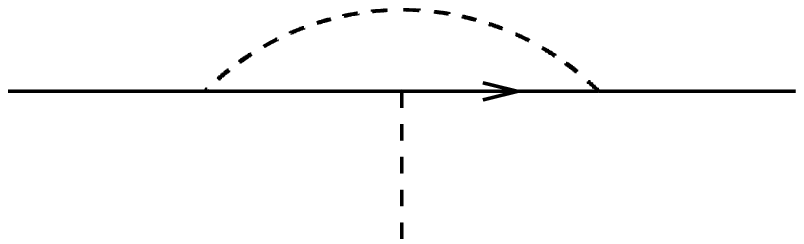}}
	
	\hspace{0.4cm}
	\subfigure[]{
		\label{fig:subfig:e}
		\includegraphics[width=2cm]{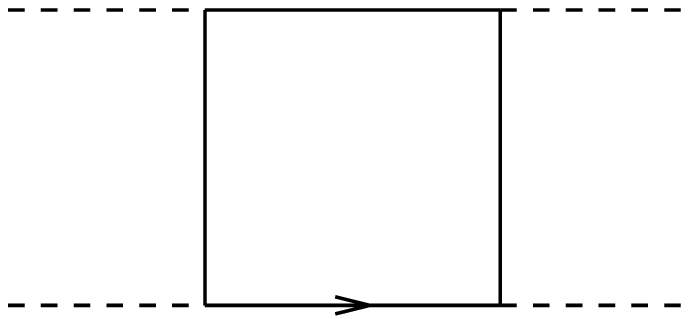}}
	\hspace{0.4cm}
	\subfigure[]{
		\label{fig:subfig:f}
		\includegraphics[width=2cm]{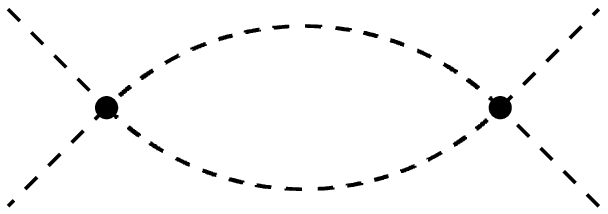}}
	\caption{One-loop graphs}
\label{fig:1l}
\end{figure}

The quantities $r_A$ depend in general on the choice of 
the Wilsonian
flow; in Table \ref{tab:1l} we give the values of $r_A$ and $c_A^{NDR}$
for the flow $n=2$, figure \ref{fig:1l} gives the corresponding
graphs. To compute the coefficients in Table \ref{tab:1l} 
one has to make an expansion in masses and momenta of graphs in
fig.\ref{fig:1l},
as mentioned after eq.(\ref{last})  

\begin{table} 
\begin{center}      
\begin{tabular}{|c||c|c|c||c|}
\hline
\rule{0pt}{3ex}structure &$c^{NDR}_{(1)}$ &$r^{\#}_{(1)}$ & $\mu \partial_{\mu}g$& graphs \\
\hline
&&&&\\$Y_{ij}$
& $-\frac{4}{\epsilon}$
& $\frac{26}{15}$
& $4$
& Fig.\ref{fig:subfig:a}\\ 
&&&&\\
$h_{ijkl}$
&$\frac{1}{\epsilon}$ 
&$-\frac{7}{12}$
&$-1$
& Fig.\ref{fig:subfig:b}\\
&&&&\\
$y_i^{\dagger} y_i$
& $- \frac{1}{\epsilon}$
&$\frac{7}{12}$
& $1$
&Fig.\ref{fig:subfig:c}\\ 
&&&&\\
$i\,y_j y_i^{\dagger} y_j$ 
& $\frac{2}{\epsilon}$
&$- \frac{47}{60}$
& $-2$
&Fig.\ref{fig:subfig:d}\\
&&&&\\
$Y_{(ijkl)}$
& $-\frac{48}{\epsilon}$
&$\frac{454}{35}$
& $48$
&Fig.\ref{fig:subfig:e}\\
&&&&\\
$h_{(ij}^{rs} h^{rs}_{kl)}$
& $\frac{3}{\epsilon}$ 
&$-\frac{7}{4}$
& $-3$
&Fig.\ref{fig:subfig:f}\\
&&&&\\
\hline
\end{tabular}
\end{center}
\caption{One-loop coefficients}
\label{tab:1l}
\end{table}

The remaining part of the one-loop bare constants is 
\begin{equation}\label{cc1}\bsp
&\Delta c_{ij}^{(1)} = 
\frac{4}{3}\left(-p^2 + \frac{2}{\e}\hat{p}^2\right) Tr \,\left[P_i P_j\right] \\
&\Delta c^{(1)} = \frac{i}{\e} \hat{\ds{p}}\,\left(2 P_i
P_i - i \gc \{S_i,P_i\}\right)  \\
&\Delta c_i^{(1)} = y_j \gc P_i y_j \\&
\Delta c_{ijkl}^{(1)} = -96~ Tr\,\left[S_{(i} S_j P_k P_{l)} + 
\frac{2}{3} P_{(i} P_j P_k P_{l)}\right] 
\esp\end{equation}

In \cite{Sch} MS is applied in BMHV in the simplest Yukawa model with a
pseudoscalar, where there is no chiral
symmetry to be maintained.  Here we introduced finite
counterterms, which are exactly those needed to obtain the same
renormalized Green functions as in the MS NDR scheme.
In \cite{Sch} it was shown that the
beta function at two loop in the MS BMHV scheme differs from the one
in MS NDR scheme by a renormalization group transformation involving
finite one-loop counterterms, which are in agreement with eq.(\ref{cc1}).

\subsection{Two-loop results}

At two loops the renormalization conditions on the 
Wilsonian effective action  read 
\begin{equation}\label{rc2bb}\bsp  
&r_{ij} = r_{ij}^{\phi} p^2 + r_{ij}^{m^2,kl} m^2_{kl}
\\
&r_{ij}^{\phi} =  r_1^{\phi} h_{ikmn}h_{jkmn} +
 r_2^{\phi} Y_{ikjk} + r_3^{\phi}  Y_{ijkk}   \\
&r_{ij}^{m^2,kl} = r_1^{m^2} h_{ikmn}h_{jlmn} +
r_2^{m^2} h_{ijmn}h_{mnkl} +  \\
&\quad\quad\quad r_3^{m^2} Y_{ikjl} + r_4^{m^2} Y_{ijkl} +
r_5^{m^2} Y_{lm} h_{ijkm}  
\esp\end{equation}

The two-point fermionic term is 
\begin{equation}\label{rc2ff}
r =  y_j^{\dagger} [
r^{\psi}_1 y_i y_i^{\dagger} y_j+  
r^{\psi}_2 y_i y_j^{\dagger} y_i+ 
r^{\psi}_3 Y_{ij}~ y_i + 
r^{\psi}_4  h_{ijkk} y_i] 
\end{equation}

The two-loop fermion-fermion-scalar term is 
\begin{equation}\label{rc2fb}\bsp
r_i = &i y_k [
r^{v3}_1 Y_{jk} y_i^{\dagger} y_j+ 
 r^{v3}_2 y_j^{\dagger} y_i y_j^{\dagger}y_k +
r^{v3}_3
(y_i^{\dagger} y_j y_j^{\dagger} +y_j^{\dagger} y_j y_i^{\dagger}) y_k+\\
&r^{v3}_4
(y_i^{\dagger} y_j y_k^{\dagger} + y_j^{\dagger} y_k y_i^{\dagger})y_j+
r^{v3}_5 y_j^{\dagger} y_i y_k^{\dagger}y_j + 
 r^{v3}_6 h_{jkll}y_i^{\dagger} y_j +
r^{v3}_7  h_{ijkl}y_j^{\dagger} y_l]
\esp
\end{equation}

The two-loop quartic scalar term is 
\begin{equation}\label{rc2b4}\bsp
&r_{ijkl} =   r^{v4}_1 Y_{ninjkl} + r^{v4}_2 Y_{nijnkl}+
r^{v4}_3 Y_{nijkln} + \\
& \quad\quad\quad h_{mnij}(r^{v4}_4 Y_{mkln} + r^{v4}_5 Y_{mknl} +
r^{v4}_6 Y_{mp} h_{pnkl} +\\
&\quad\quad\quad r^{v4}_7 h_{mnpq} h_{pqkl} + r^{v4}_8 h_{mpkl} h_{npqq} +
r^{v4}_9 h_{npqk} h_{mpql})
\esp
\end{equation}
where symmetrization over the indices $i,j,k,l$ is understood.

\begin{table}
\begin{center}
\begin{tabular}{|c||c|c|c|c||c|}
\hline
\rule{0pt}{3ex}
structure&$c^{NDR}$ & $r^\f,\,r^{m^2}$ & $\mu \partial_{\mu}g$ & $\tau$ & graphs  \\
\hline
&&&&&\\
$h_{ikmn}h_{jkmn}p^2$
& $-\frac{1}{12 \epsilon}$
& $\frac{7}{432} + \frac{5 v}{81}$
& $\frac{1}{6}$
& $0$
& Fig. \ref{fig:bb:a}\\ 
&&&&&\\
$Y_{ikjk}p^2$
&$-\frac{8}{\epsilon^2}+ \frac{2}{\epsilon}$  
& $-\frac{25}{9} + \frac{178 v}{81}$
&$- \frac{164}{15}$
& $0$
&Fig. \ref{fig:bb:b}\\
&&&&&\\
$Y_{ijkk}p^2$
& $-\frac{4}{\epsilon^2}+\frac{3}{ \epsilon}$
& $-\frac{16021}{8505} + \frac{49 v}{729}$
& $- \frac{279}{35}$
& $\frac{157}{105}$ 
& Fig. \ref{fig:bb:c}\\ 
&&&&&\\
$h_{ikmn}h_{jlmn}m^2_{kl}$
& $\frac{1}{\epsilon^2}-\frac{1}{2 \epsilon}$
& $\frac{67}{108} + \frac{29 v}{324}$
& 
& 
& Fig. \ref{fig:bb:a} \\
&&&&&\\
$h_{ijmn} h_{mnkl}m^2_{kl}$
& $\frac{1}{\epsilon^2}$
& $\frac{49}{144}$
& 
&  
& Fig. \ref{fig:bb:d} \\
&&&&&\\
$Y_{ikjl}m^2_{kl}$
& $-\frac{8}{\epsilon^2}+\frac{4}{\epsilon}$
& $-\frac{1642}{729} + \frac{5078 v}{2187}$
& 
&   
& Fig. \ref{fig:bb:b}   \\
&&&&&\\
$Y_{ijkl}m^2_{kl}$
& $-\frac{16}{\epsilon^2}$
& $-\frac{146}{81} + \frac{4 v}{243}$
& 
&  
& Fig. \ref{fig:bb:c}    \\
&&&&&\\
$h_{ijkm} Y_{lm}m^2_{kl}$
& $\frac{4}{\epsilon^2}-\frac{2}{\epsilon}$ 
& $\frac{4421}{545} + \frac{121 v}{27}$
& 
&  
& Fig. \ref{fig:bb:e} \\
&&&&&\\
\hline
\end{tabular}
\end{center}
\caption{Two-loop coefficients for the two-point bosonic functions}
\label{tab:2l1}
\end{table}

\begin{figure} 
\centering
\subfigure[]{\label{fig:bb:a}\includegraphics[width=2cm]{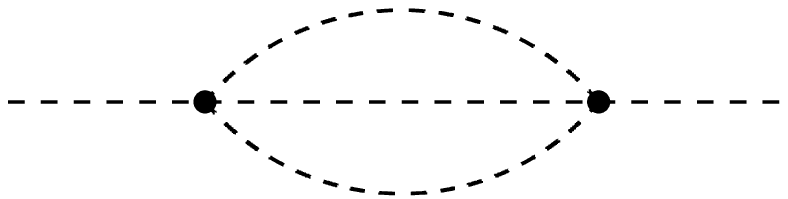}}
	\hspace{0.4cm}
\subfigure[]{\label{fig:bb:b}\includegraphics[width=2cm]{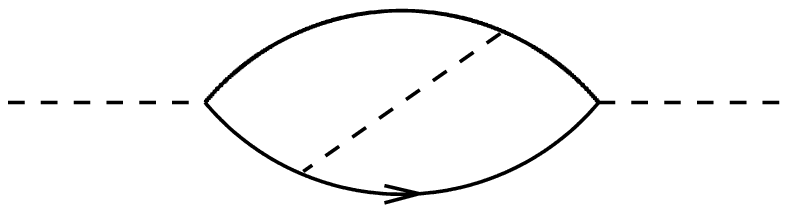}}
	\hspace{0.4cm}
\subfigure[]{\label{fig:bb:c}\includegraphics[width=2cm]{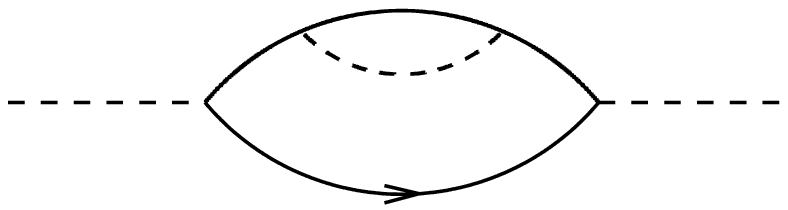}}
	\hspace{0.4cm}
\subfigure[]{\label{fig:bb:d}\includegraphics[width=2cm]{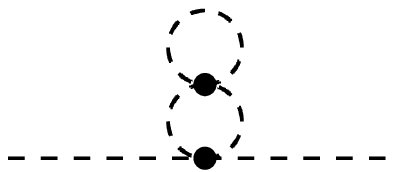}}
	\hspace{0.4cm}
\subfigure[]{\label{fig:bb:e}\includegraphics[width=2cm]{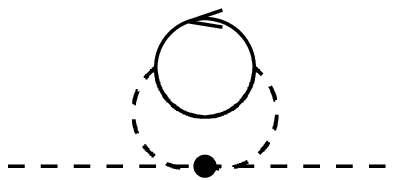}}
\caption{Two-loop graphs for the two-point bosonic function}
\label{fig:2l1}
\end{figure}

\begin{table}
\centering
\begin{tabular}{|c||c|c|c|c||c|}
\hline
\rule{0pt}{3ex}
structure & $c^{NDR}$ &$r^\psi$ & $\mu \partial_{\mu}g$ & $\tau$ 
& graphs \\
\hline
&&&&& \\
$y_j^{\dagger} y_i y_i^{\dagger} y_j$
& $-\frac{1}{2 \epsilon^2} + \frac{1}{8 \epsilon}$
& $-\frac{2359}{9720} + \frac{289 v}{5832}$
& $-\frac{29}{40}$
& $\frac{13}{120}$
& Fig. \ref{fig:ff:a}\\ 
&&&&& \\
$y_j^{\dagger} y_i y_j^{\dagger} y_i$
&$-\frac{2}{\epsilon^2}$ 
& $-\frac{229}{324} + \frac{319 v}{486}$
&$-\frac{7}{3}$
& $0$ 
& Fig. \ref{fig:ff:b} \\
&&&&& \\
$y_j^{\dagger} y_i Y_{ij}$
& $-\frac{2}{\epsilon^2}+\frac{3}{2 \epsilon}$
& $- \frac{2375}{486} - \frac{6107 v}{1458}$
& $- \frac{127}{30}$
& $\frac{11}{10}$ 
& Fig. \ref{fig:ff:c}  \\ 
&&&&& \\ 
$h_{ijkk} y_j^{\dagger} y_i$
& $0$ 
& $\frac{13}{30}$
& $0$
& $0$
& Fig. \ref{fig:ff:d}  \\ 
&&&&& \\ 
\hline
\end{tabular}
\caption{Two-loop coefficients for the fermionic two-point functions  }
\label{tab:2l2}
\end{table}

\begin{figure} 
\centering
\subfigure[]{\label{fig:ff:a}\includegraphics[width=2cm]{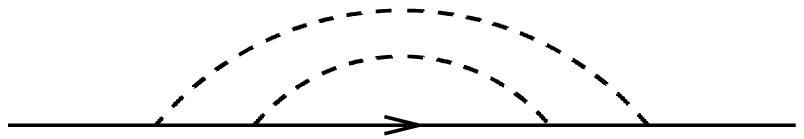}}
	\hspace{0.4cm}
\subfigure[]{\label{fig:ff:b}\includegraphics[width=2cm]{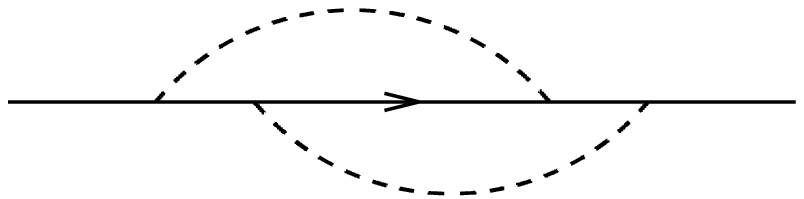}}
	\hspace{0.4cm}
\subfigure[]{\label{fig:ff:c}\includegraphics[width=2cm]{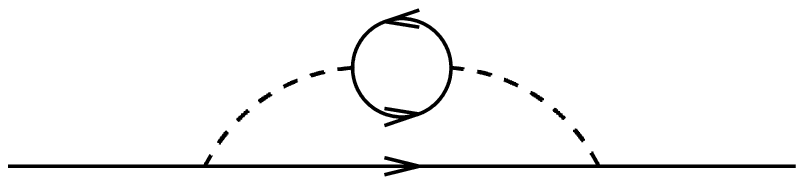}}
	\hspace{0.4cm}
\subfigure[]{\label{fig:ff:d}\includegraphics[width=2cm]{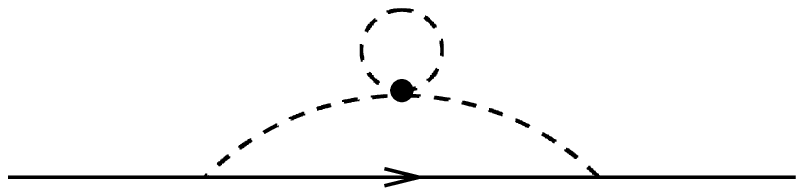}}
\caption{Two-loop graphs for the two-point fermionic function}
\label{fig:2l2}
\end{figure}

\begin{table}
\centering
\begin{tabular}{|c||c|c|c|c||c|}
\hline 
\rule{0pt}{3ex} 
structure &  $c^{NDR}$ &$r^{v3}$ & $\mu \partial_{\mu}g$ & $\tau$ 
 & graphs \\
\hline 
&&&&& \\ 
$Y_{jk} y_k y_i^{\dagger} y_j$
& $\frac{4}{\epsilon^2}-\frac{2}{\epsilon}$
& $\frac{19384}{2835} + \frac{197 v}{243}$
& $\frac{647}{105}$
& $- \frac{34}{35}$
& Fig. \ref{fig:v3:a} \\ 
&&&&& \\ 
$y_k y_j^{\dagger} y_i y_j^{\dagger}y_k$
&$\frac{2}{\epsilon^2}-\frac{1}{\epsilon}$
& $\frac{47}{81} + \frac{13 v}{486}$
&$\frac{107}{30}$
& $0$ 
& Fig. \ref{fig:v3:b} \\
&&&&& \\ 
$y_k(y_i^{\dagger} y_j y_j^{\dagger} +y_j^{\dagger} y_j y_i^{\dagger})
y_k$
& $\frac{1}{\epsilon^2}-\frac{1}{2 \epsilon}$
& $\frac{10009}{22680} - \frac{83 v}{972}$
& $\frac{647}{420}$
& $-\frac{17}{70}$
&  Fig. \ref{fig:v3:c}\\
&&&&& \\ 
$y_k (y_i^{\dagger} y_j y_k^{\dagger} + y_j^{\dagger} y_k
y_i^{\dagger})y_j$
& $\frac{2}{\epsilon^2}$ 
& $\frac{415}{972} - \frac{169 v}{1458}$
& $\frac{47}{30}$ 
& $0$ 
&  Fig. \ref{fig:v3:d} \\
&&&&& \\
$y_k y_j^{\dagger} y_i y_k^{\dagger}y_j$
& $ \frac{1}{\epsilon}$
& $\frac{76}{729} - \frac{920 v}{2187}$
& $- 2$
& $0$
&  Fig. \ref{fig:v3:e}\\
&&&&& \\
$h_{jkll}y_k y_i^{\dagger} y_j$
& $0$ 
& $- \frac{33}{70}$
& $0$
& $0$
&  Fig. \ref{fig:v3:f} \\
&&&&& \\
$h_{ijkl}y_k y_j^{\dagger} y_l$
& $- \frac{1}{\epsilon}$
& $\frac{7}{81} + \frac{160 v}{243}$
& $2$
& $0$
&  Fig. \ref{fig:v3:g}\\
&&&&& \\ 
\hline
\end{tabular}
\caption{Two-loop coefficients for the cubic vertex  }
\label{tab:2l3}
\end{table}

\begin{figure} 
\centering
\subfigure[]{\label{fig:v3:a}\includegraphics[width=2cm]{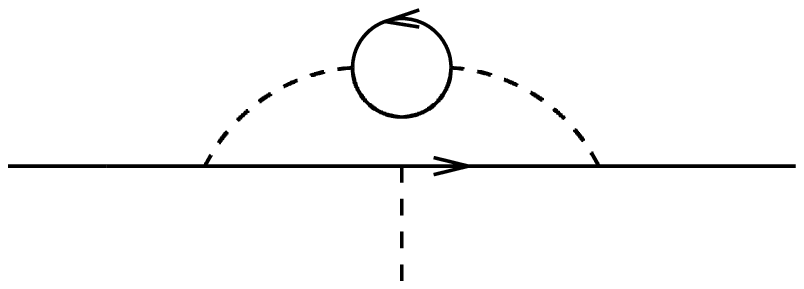}}
	\hspace{0.4cm}
\subfigure[]{\label{fig:v3:b}\includegraphics[width=2cm]{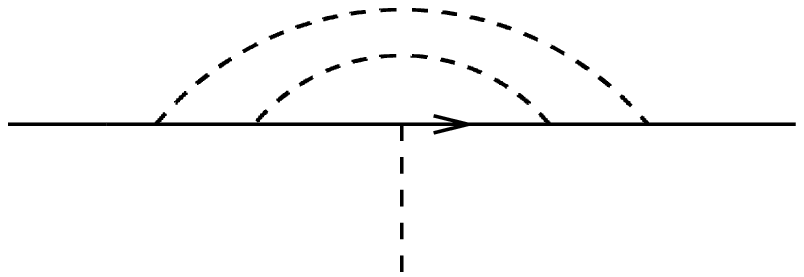}}
	\hspace{0.4cm}
\subfigure[]{\label{fig:v3:c}\includegraphics[width=2cm]{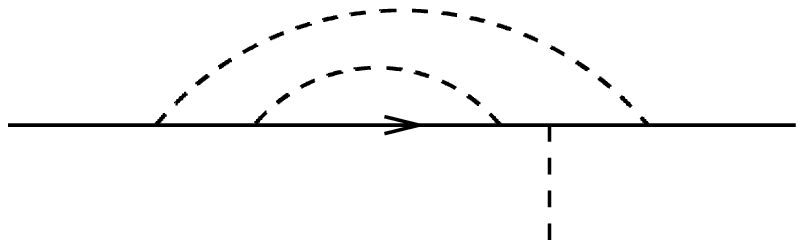}}
	\hspace{0.4cm}
\subfigure[]{\label{fig:v3:d}\includegraphics[width=2cm]{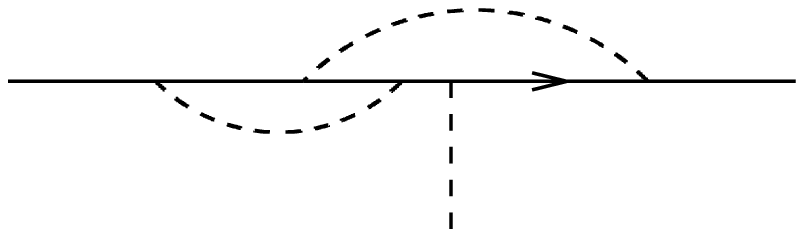}}
	\hspace{0.4cm}
\subfigure[]{\label{fig:v3:e}\includegraphics[width=2cm]{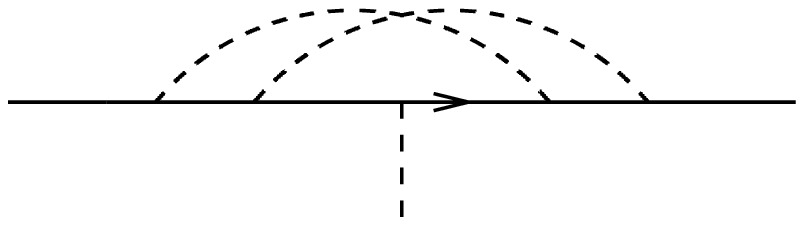}}
	\hspace{0.4cm}
\subfigure[]{\label{fig:v3:f}\includegraphics[width=2cm]{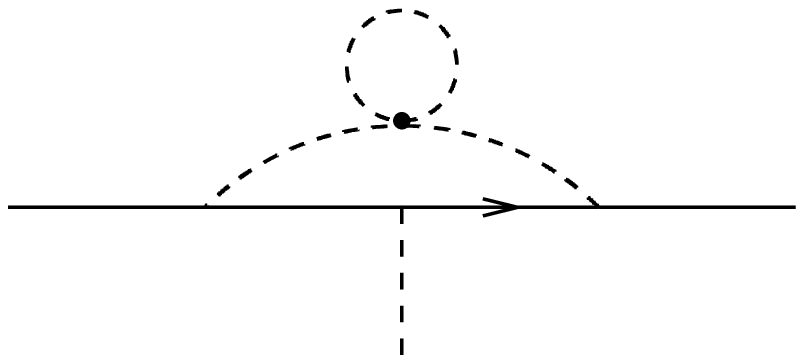}}
	\hspace{0.4cm}
\subfigure[]{\label{fig:v3:g}\includegraphics[width=2cm]{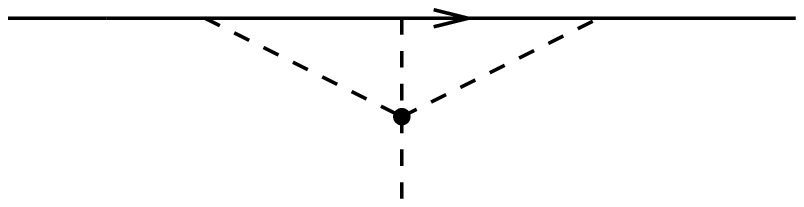}}
\caption{Two-loop graphs for the cubic vertex}
\label{fig:2l3}
\end{figure}

\begin{table}
\centering
\begin{tabular}{|c||c|c|c|c||c|}
\hline 
\rule{0pt}{3ex}
 structure & $c^{NDR}$ & $r^{v4}$ & $\mu \partial_{\mu}g$ & $\tau$ 
& graphs \\
\hline 
&&&&& \\
$Y_{ninjkl}$
& $-\frac{192}{\epsilon^2}+\frac{96}{\epsilon}$
& $ -\frac{3782672}{76545}+\frac{66416 v}{6561}$
& $-\frac{10352}{35}$
& $0$
& Fig. \ref{fig:v4:a}\\ 
&&&&& \\ 
$Y_{nijnkl}$
& $\frac{48}{\epsilon}$
& $-\frac{15088}{18225} + \frac{26272 v}{2187}$
& $-96$
& $0$
& Fig. \ref{fig:v4:b}\\
&&&&& \\ 
 $Y_{nijkln}$
& $-\frac{96}{\epsilon^2}+\frac{48}{\epsilon}$
& $-\frac{5660407}{91854} + \frac{155224 v}{19683}$
& $-\frac{13756}{105}$
& $\frac{1772}{105}$
& Fig. \ref{fig:v4:c}  \\ 
&&&&& \\ 
$h_{mnij} Y_{mkln}$
& $-\frac{96}{\epsilon^2}$
& $-\frac{292}{27} + \frac{8 v}{81}$
& $-112$ 
& $0$
& Fig. \ref{fig:v4:d}\\ 
&&&&& \\
$h_{mnij} Y_{mknl}$
& $-\frac{48}{\epsilon^2}+\frac{24}{\epsilon}$
& $-\frac{4256}{243} + \frac{10156 v}{729}$
& $-104$
& $0$ 
&  Fig. \ref{fig:v4:e}\\ 
&&&&& \\
$h_{mnij} Y_{mp} h_{pnkl} $
& $\frac{12}{\epsilon^2}-\frac{6}{\epsilon}$
& $\frac{949}{36} + \frac{121 v}{9}$
& $\frac{107}{5}$
& $-\frac{23}{5}$
& Fig. \ref{fig:v4:f} \\ 
&&&&& \\
$h_{mnij} h_{mnpq} h_{pqkl}$
& $\frac{3}{\epsilon^2}$
& $\frac{61}{48}$
& $\frac{7}{2}$
& $0$
& Fig. \ref{fig:v4:g}\\ 
&&&&& \\
$h_{mnij} h_{mpkl} h_{npqq}$
& $0$
& $-\frac{39}{20}$
& $0$
& $0$
& Fig. \ref{fig:v4:h} \\ 
&&&&& \\
$h_{mnij} h_{npqk} h_{mpql}$
& $\frac{6}{\epsilon^2} - \frac{3}{\epsilon}$
& $\frac{38}{9}+\frac{29 v}{54} $
& $13$
& $0$
& Fig. \ref{fig:v4:i} \\ 
&&&&& \\
\hline
\end{tabular}
\caption{Two-loop coefficients for the quartic vertex}
\label{tab:2l4}
\end{table}

\begin{figure} 
\centering
\subfigure[]{\label{fig:v4:a}\includegraphics[width=2cm]{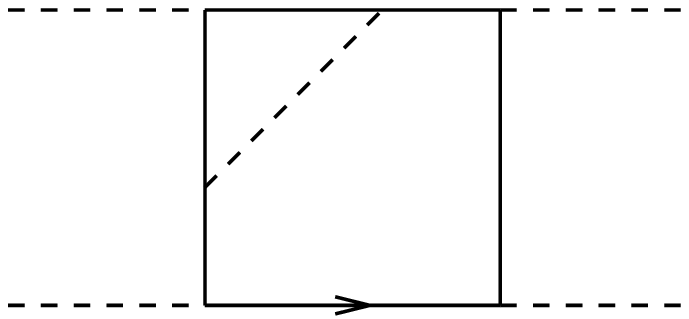}}
	\hspace{0.4cm}
\subfigure[]{\label{fig:v4:b}\includegraphics[width=2cm]{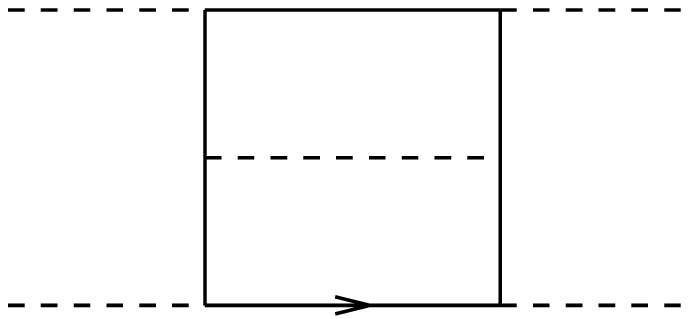}}
	\hspace{0.4cm}
\subfigure[]{\label{fig:v4:c}\includegraphics[width=2cm]{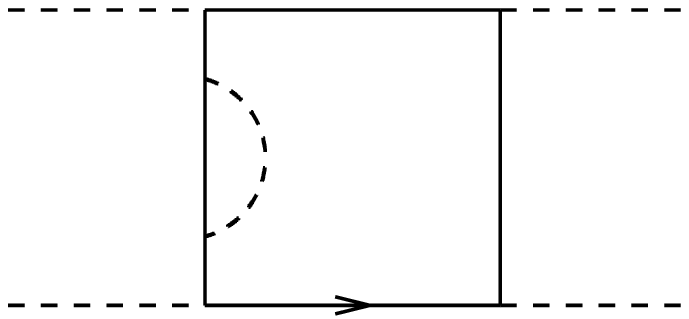}}
	\hspace{0.4cm}
\subfigure[]{\label{fig:v4:d}\includegraphics[width=2cm]{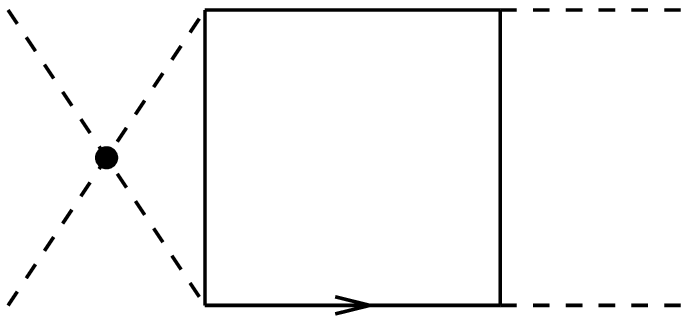}}
	\hspace{0.4cm}
\subfigure[]{\label{fig:v4:e}\includegraphics[width=2cm]{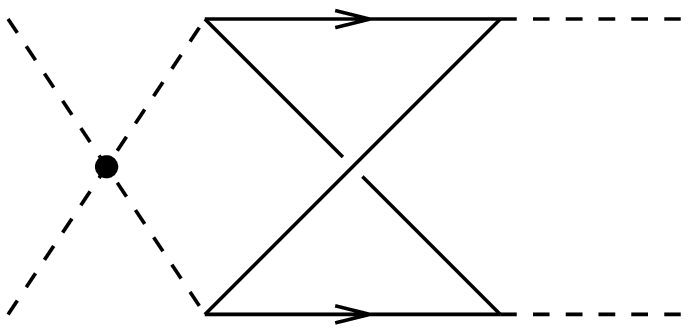}}
	\hspace{0.4cm}
\subfigure[]{\label{fig:v4:f}\includegraphics[width=2cm]{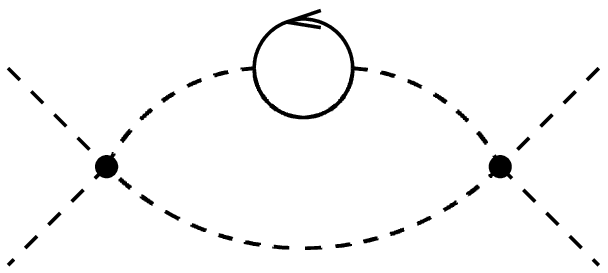}}
	\hspace{0.4cm}
\subfigure[]{\label{fig:v4:g}\includegraphics[width=2cm]{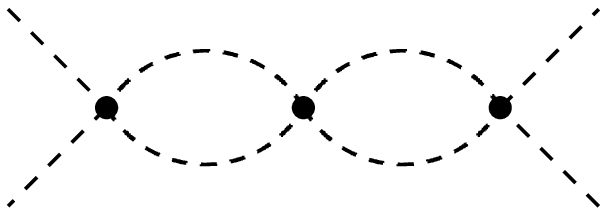}}
	\hspace{0.4cm}
\subfigure[]{\label{fig:v4:h}\includegraphics[width=2cm]{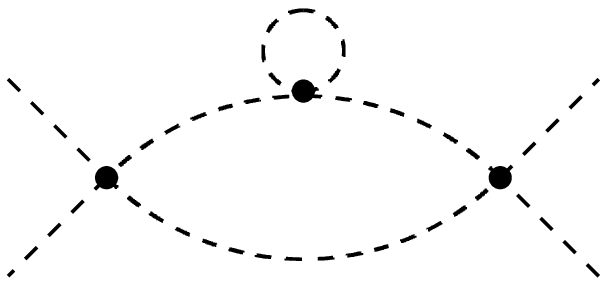}}
	\hspace{0.4cm}
\subfigure[]{\label{fig:v4:i}\includegraphics[width=2cm]{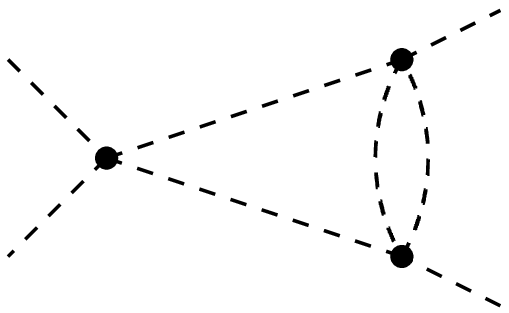}}
\caption{Two-loop graphs for the quartic vertex}
\label{fig:2l4}
\end{figure}

The coefficients in eqs.(\ref{rc2bb}), (\ref{rc2ff}), (\ref{rc2fb}), (\ref{rc2b4}) and the
corresponding ones for the two-loop $c_A^{NDR}$ in the case of the 
Wilsonian flow $n=2$
are given respectively in the Tables \ref{tab:2l1}, \ref{tab:2l2},
\ref{tab:2l3}, \ref{tab:2l4}. 
Figures \ref{fig:2l1}, \ref{fig:2l2}, \ref{fig:2l3}, \ref{fig:2l4}
 show the graphs
contributing
to these coefficients; for each diagram shown there are in general 
corresponding graphs with one-loop counterterms that are not represented.

The naive part $c^{NDR}_A$ of the two-loop bare constants agree with
\cite{Mac}.
In order to make the comparison with \cite{Mac}, where a two-component
spinor formalism is used, one must substitute the traces 
$k \,Tr\left[  Y_{i_1} Y^{\dagger}_{i_2} ... Y_{i_{2n-1}}
Y^{\dagger}_{i_{2n}}\right]$ in \cite{Mac}
with $Y_{i_1i_2...i_{2n-1}i_{2n}}$.

The remaining part of the bare counterterms is given in the Appendix.

With the bare action determined using the flow  $n=2$, 
according to the lines of sect.(\ref{subsec:wilseff})   one can define a new Wilsonian action $\Gamma^{(n=3)}_\L$.  For $n=3$ (or bigger) the
new effective action is finite and satisfies the Ward identities. The graphs 
of  $\Gamma^{(n=3)}_\L$ now have  $n=3$ propagators $D_{\L}$ but contain
the  $n=2$ counterterms. This suggests the following consistency check:  
consider the contributions of $n=3$ bare (marginal) graph, of its relative countergraph
and two loop counterterm; we have verified  that although  the two
first quantities are divergent and separately 
not invariant, the total sum is finite and invariant.
Furthermore this total sum, being computed on a different flow is
different from the $n=2$ case.

\subsection{Renormalization group equation}\label{sec:rge}

Let us compute the renormalization group beta and gamma functions in
the BMHV scheme at two loops.
In the case of the simplest Yukawa model this has been
done in \cite{Sch} using the minimal subtraction formulas \cite{tH}
and taking into account the one-loop evanescent tensors in the bare
action; after obtaining a renormalization group equation involving
also insertions of evanescent tensors, these are solved in terms of
relevant
couplings obtaining the usual renormalization group equation.

To avoid making a similar subtle analysis of the bare couplings in our
case,
we will obtain the beta and gamma functions working directly with the
Wilsonian effective action, which satisfies the effective 
renormalization group equation \cite{PR}.
We will restrict to the massless case; in \cite{PR} it is considered
also the massive case.

In the compact notation of subsection \ref{subsec:wilseff}
the Gell-Mann and Low renormalization group equation reads 
\begin{equation}\label{rge}
(\m \frac{\pa }{\pa \mu }+
\beta \cdot \frac{\pa}{\partial g } + 
J \g_{\f}^T \frac{\d}{ \d J }) Z[J] = {\cal{E}}[J]
\end{equation}
where $Z$ is the renormalized functional in which  the limit
$\e \to 0$ has not yet been taken;
${\cal{E}}$ is an evanescent functional, which in general is not
vanishing for finite value of $\e$ since the renormalization of 
the theory  is not strictly
multiplicative, and
\begin{equation}
\beta \cdot \frac{\partial}{\partial g } =
\beta_{ijkl}  \frac{\partial}{\partial h_{ijkl} } +
Tr (\beta_{Y_i} \frac{\partial}{\partial Y_i} +
\beta_{Y_i}^{\dagger} \frac{\partial}{\partial Y_i^{\dagger}}) 
\end{equation}

From eq.(\ref{ZL}) it follows that $Z_{\L}[J]$
satisfies the `effective renormalization group equation' 
\begin{equation}\label{rgef}
(\mu \frac{\partial}{\partial \mu }+ 
\beta \cdot \frac{\partial}{\partial g } + 
J \g_{\phi}^T \frac{\d}{\d J }
-\hbar\frac{\d }{\d J} \g_{\phi}^T D_{\L}^{-1} K_{\L}  
\frac{\d}{\d J} ) Z_{\L}[J]
= {\cal{E}}_{\L}[J]
\end{equation}

Define the functional $Z_{\L}[J,\chi]$ as in (\ref{Zj}), but with 
\begin{equation}\label{break}
S_{\L}(\Phi,\chi) = S_{\L}(\Phi) - \chi \Delta_{\g} [\Phi] \quad;\quad
 \Delta_{\g} [\Phi] \equiv \Phi \g_{\phi}^T D_{\L}^{-1} K_{\L}  \Phi
\end{equation}
The effective renormalization group equation can be written as
\begin{equation}
(\mu \frac{\partial}{ \partial \mu }+ 
\beta \cdot \frac{\partial}{\partial g } + 
J \g_{\phi}^T \frac{\d}{ \d J } 
- \frac{\partial}{\partial \chi } )|_{\chi = 0} Z_{\L}[J,\chi] = 
{\cal{E}}_{\L}[J] 
\end{equation}
which in terms of the $1$-PI functional generator reads
\begin{eqnarray}
(\mu \frac{\partial}{ \partial \mu }+ 
\beta \cdot \frac{\partial}{ \partial g } -
\Phi \g_{\phi} \frac{\d}{ \d \Phi }
- \frac{\partial}{ \partial \chi } )|_{\chi = 0}) 
\Gamma_{\L}[\Phi,\chi] = {\cal{E}}_{\L}[\Phi] 
\end{eqnarray}
or equivalently 
\begin{eqnarray}\label{ERG}
(\mu \frac{\partial}{ \partial \mu }+ 
\beta \cdot \frac{\partial}{ \partial g } -
\Phi \g_{\phi} \frac{\d}{\d \Phi }) \Gamma_{\L} [\Phi] =
 {\cal T}^{\g}_{\L}[\Phi]+ {\cal{E}}_{\L}[\Phi]
\end{eqnarray}
where ${\cal T}^{\g}_{\L}[\Phi]$ represents the insertion 
of the non local operator $\Delta_{\g} [\Phi]$ on the functional $\Gamma_{\L} [\Phi]$; its Feynman graphs
contain the new vertices

\begin{minipage}[b]{0.25\textwidth}
\rule{2cm}{0pt}\raisebox{1.1ex}[0cm][0cm]{\includegraphics[width=2cm]{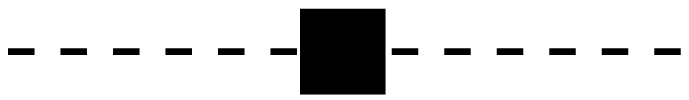}}\\
\rule{2cm}{0pt}\raisebox{.2ex}[0cm][0cm]{\includegraphics[width=2cm]{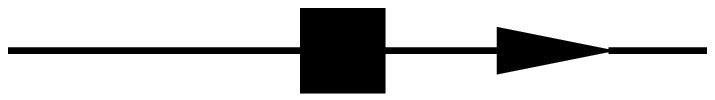}}
\end{minipage}
\begin{minipage}[b]{0.7\textwidth}
\begin{equation}\label{ins}
\bsp 
&2 K_{\L}(p) D_H^{-1}(p) \g_{ij}\\
&2 K_{\L}(p) S_H^{-1}(p)\g_{\psi}
\esp
\end{equation}
\end{minipage}
between two bosonic or fermionic lines; 
due to its nonlocality $\Delta_{\g}[\Phi]$ does not require 
renormalization.

Let us finally rewrite in a  suitable way the gradient terms in (\ref{ERG});
using (\ref{res}) one gets 
\begin{eqnarray}\label{grad}
\mu \frac{\partial g_A^{(1)}}{\partial \mu} = g_{A,-1}^{(1,0)} =
- c_{A,-1}^{(1)} ~~;~~
\mu \frac{\partial g_A^{(2)}}{\partial \mu} = 2 g_{A,-1}^{(2,0)} +
g_{A,-1}^{(1,1)} 
\end{eqnarray}
These gradients have the same group structure as the renormalization
constants given in eqs.(\ref{rc1}), (\ref{rc2bb});
the coefficients of these group structures
 are given in the tables
\ref{tab:1l}, \ref{tab:2l1}, \ref{tab:2l2}, \ref{tab:2l3},
\ref{tab:2l4} in the case of the 
$n=2$ flow. Observe that, while $g_{A,-1}^{(2,0)}$ and 
$g_{A,-1}^{(1,1)}$ are not separately chiral group-covariant, the
gradient term $2 g_{A,-1}^{(2,0)} +g_{A,-1}^{(1,1)}$ must be covariant.
This provides a check on the computations, besides the double pole
rule mentioned after (\ref{cons}), that we made in a systematic way.

At one loop the Wilsonian gradients can be expressed 
in terms of the pole of the bare couplings, so that the
formulae for the beta and gamma functions are equivalent to the
standard formulae \cite{tH}; using $\beta_A = \sum \frac{\hbar^l}{(4
\pi)^{2l}} \beta_A^{(l)} $ and defining $ \beta_i = \beta_{Y_i} P_R +
\beta_{Y_i}^\dagger P_L $
we get 
\begin{eqnarray}\label{beta1}
&&\g_{ij}^{(1)} = 2 Y_{ij} ~~;~~
\g^{\bar\psi (1)} =\frac{1}{2} y_i y_i^{\dagger} ~~;~~ 
\g^{\psi (1)} =\frac{1}{2} y_i^{\dagger} y_i\nonumber \\
&&\beta_i^{(1)} = 2 y_j y_i^{\dagger} y_j + 
\frac{1}{2}(y_i y_j^{\dagger} y_j + y_j y_j^{\dagger} y_i) +
2 Y_{ij} y_j \\
&&\beta^{(1)}_{ijkl} = -48 Y_{(ijkl)}
+ 3 h^{rs}_{(ij} h^{rs}_{kl)} + 8 Y_{m(i} h^m_{~jkl)}\nonumber 
\end{eqnarray}

At two loops the Wilsonian gradients are not given by the simple pole
of the bare counterterms (\ref{grad});  
the beta and gamma functions are 
given in terms of these Wilsonian gradients, the ${\cal{T}}$
insertions, given in the tables
\ref{tab:2l1}, \ref{tab:2l2}, \ref{tab:2l3},
\ref{tab:2l4},
and the one-loop renormalization conditions (\ref{rc1}). 

We get
\begin{eqnarray}\label{beta2}
&&\g_{ij}^{(2)} =
\frac{1}{12} h_{ikmn} h_{jkmn} - 2 Y_{ikjk} - 3 Y_{ijkk}  \nonumber \\
&&\g^{\psi (2)} = 
- \frac{1}{8} y^{\dagger}_j y_i y^{\dagger}_i y_j -
\frac{3}{2} Y_{ij} y^{\dagger}_j y_i
 \nonumber \\
&&\g^{\bar\psi (2)} = 
- \frac{1}{8} y_j y_i^{\dagger} y_i y_j^{\dagger} -
\frac{3}{2} Y_{ij} y_j y_i^{\dagger}\\
&&\beta_i^{(2)} = 
2 y_k y_j^{\dagger}y_i (-y_j^{\dagger}y_k+y_k^{\dagger}y_j) -
2 h_{ijkl} y_j y_k^{\dagger}y_l -
4 Y_{jk} y_j y^{\dagger}_i y_k -  \nonumber  \\
&&\qquad\quad y_k (y^{\dagger}_i y_j y^{\dagger}_j + y^{\dagger}_j y_j
y^{\dagger}_i) y_k + y_i \g^{\psi (2)} + \g^{\bar\psi (2)}y_i +
\g_{ij}^{(2)} y_j
 \nonumber \\
&&\beta^{(2)}_{ijkl} = 
96 (2 Y_{ninjkl} + Y_{nijnkl} + Y_{nijkln}) + \nonumber \\
&&\qquad\quad 6 h_{mnij}(8 Y_{mknl} - 2 Y_{mp} h_{pnkl} -  h_{npqk}
h_{mpql})+ 4 \g_{im}^{(2)} h_{mjkl}
\nonumber
\end{eqnarray}

where in the last equation the indices $i,j,k,l$ must be totally symmetrized.

These formulae are in agreement with \cite{Mac}. No
renormalization group transformation is needed, because the
renormalization conditions are the same.

As an example, we give the separate contributions to the term
$Y_{nijkln}$ of $\beta^{(2)}_{ijkl}$.
In (\ref{ins}) are indicated the Feynman rules for the insertions
${\cal T}_\L^\g $.

$-\mu \frac{\partial g_{ijkl}^{(2)}}{\partial \mu}$ (see table \ref{tab:2l4})
contributes $\frac{13756}{105}$.

$-\beta^{(1)}\cdot \frac{\pa}{\pa g} g_{ijkl}^{(1)}$ gets the
contribution $-\frac{1816}{35}$ due to
the term $\frac{454}{35} ~Y_{(ijkl)}$  of
$g_{ijkl}^{(1)}$ (see table \ref{tab:1l}) and the term 
$\frac{1}{2}(y_i y_j^{\dagger} y_j + y_j y_j^{\dagger} y_i)$
of $\beta^{(1)}_i$.

\begin{figure}
\centering
\includegraphics[width=3cm]{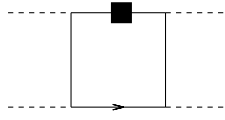}
\caption{$4$-point ${\cal T}$  Green function contributing to $Y_{nijkln}$}
\label{fig:tau}
\end{figure}

Finally $ {\cal T}_\L^\g $  
contributes to $Y_{nijkln}$ through the graph in fig. \ref{fig:tau}
and gives the coefficient $\frac{1772}{105}$.

These three contributions add up to the expected value  
$96 ~Y_{nijkln}$ for  $\beta^{(2)}_{ijkl}$ 
in equation (\ref{beta2}).

\section{External currents and local Ward identities}
\subsection{External currents}

In order to define the currents associated to the symmetry
transformations
(\ref{group}) we will
add in the action (\ref{bareY}) an external gauge field $A_{\mu}^a$.

At tree level the action is now
\begin{equation}\label{treeact}
S^{(0)} = \int \bar{\psi} \g_\m D_\m \psi +
\oh (D_\m \phi)^2 + i \bar{\psi} y_i \psi \phi_i +  
\frac{1}{4!} h_{ijkl} \phi_{i} \phi_{j} \phi_{k} \phi_{l}
\end{equation}
where
\begin{equation}
D_\m \psi = (\pa_\m + i A_\mu^a t^a) \psi \quad;\quad
D_\m \phi_i = (\pa_\m \d_{ij} + 
i A_\mu^a \theta^a_{ij}) \phi_j 
\end{equation}

We can choose for the external gauge field  the condition 
$\hat{A}_{\mu}^a = 0$, so that $\g_{\mu} t^a$ is equivalent to
$\bar{t}^a \g_{\mu}$ and the fermionic gauge coupling 
$i\g_{\mu} t^a$  
is  equal to the reflection-symmetric expression
$\frac{i}{2}(\g_{\mu} t^a + \bar{t}^a \g_{\mu})$.

We set the $\mu^\e$ factors to one;
they can be easily reintroduced by dimensional analysis.

The local  infinitesimal version of transformations (\ref{group}) is now
\begin{eqnarray}\label{group4}
&&\d A_{\mu}^a = - \partial_{\mu} \e^a - 
f^{abc} A_{\mu}^c \e^b = - D_\m^{ab} \e^b \no \\
&&\d \psi = i \e^a t^a \psi ~~;~~
\d \bar{\psi} = -i \e^a \bar{\psi} \bar{t}^a   \\
&&\d \phi_i = i \e^a \theta^a_{ij} \phi_j \no
\end{eqnarray}
where $\hat{\partial}_{\mu} \e^a = 0$.
The tree-level action is invariant under these
transformations apart from an evanescent fermionic kinetic
term.

To renormalize arbitrary products of currents we will consider in the
bare action local monomials in $A_\mu^a$ up to fourth order.

The bare action (\ref{bareY}) gets additional terms: 
\begin{equation}\label{sa}\bsp
S^{(A)} = \int &\oh c_{\m \n}^{ab}  A_\mu^a A_\n^b +
\frac{1}{3!} c_{\m \n \r}^{abc}
A_\mu^a A_\n^b A_\r^c + \frac{1}{4!} c_{\m \n \r \s}^{abcd}
 A_\mu^a A_\n^b A_\r^c A_\s^d +  \\
&A_\mu^a \bar{\psi} c_\mu^a \psi + 
\frac{1}{2} c_{\m ij}^a A_\mu^a \phi_i \phi_j
+ \frac{1}{4}c_{\m \n ij}^{ab} A_\mu^a A_\n^b \phi_i \phi_j  
\esp\end{equation}

The marginal part $S_W$ of the Wilsonian effective action  will have a similar
dependence on $A_\mu^a$.

The renormalization conditions on $S_W$ analogous to (\ref{grf}),
which include the new vertices, have
to be chosen in such
a way that the local 
chiral Ward identities hold; to this aim the renormalization
conditions 
are fixed compatibly with the
effective Ward identities. These identities have been
introduced in \cite{Becchi} and further studied in \cite{PRR}.

 An analysis similar to the one in sect.\ref{sec:rge}
shows that the
extra term of the effective Ward identity can be represented as the insertion of 
non-local vertices. 
Under transformation (\ref{group4}) one gets:
\begin{equation}\label{deltag}
\d \Gamma^{\L} = {\cal{T}}^{\L} + 
{\cal O}^{\L} \qquad;\qquad
{\cal T}^{\L} = \left[\d\left(\oh\Phi D_{\L}^{-1}K_{\L}
\Phi\right)\right]\cdot \Gamma^{\L}
\end{equation}
where ${\cal{T}}^{\L}$ is the generator of the $1$-PI Green
functions with one insertion of the operator $\d\left(\oh\Phi D_{\L}^{-1}K_{\L}
\Phi\right)$ . Its  Feynman rules are
\begin{eqnarray}\label{inswi1}
-i \e^a(p) [S_H^{-1}(p+q) K_{\L}(p+q) t^a-
\bar{t}^a S_H^{-1}(q) K_{\L}(q)]
\end{eqnarray}
for insertion on the fermionic lines and
\begin{eqnarray}\label{inswi2}
-i \e^a(p) [D_H^{-1}(p+q) K_{\L}(p+q) \theta^a-
\theta^a D_H^{-1}(q) K_{\L}(q)]
\end{eqnarray}
for insertion on the scalar lines.

If one can choose the renormalization conditions on $\Gamma^{\L}$
such that $\cal{O}^{\L}$ is evanescent, the effective Ward
identities
are not anomalous and the usual Ward identities are recovered 
at $\L = 0$; otherwise it is anyway possible to separate the functional
$\cal{O}^{\L}$ in a part corresponding to the local anomaly operator
plus a part representing an evanescent
operator:
\begin{eqnarray}
\cal{O}^{\L} = \cal A^{\L} + \cal{E}^{\L}
\end{eqnarray}
Equation (\ref{deltag}) is the effective Ward identity 
\cite{Becchi,PRR}, cast in a
form which we find convenient for explicit computations.

The anomaly operator is defined by its renormalization conditions at
the
renormalization scale $\L = \m$ therefore we have to consider the
marginal 
projection of eq.(\ref{deltag}) in the limit $\e \to 0$:
\begin{eqnarray}\label{ewi}
\d S_W = {\cal{T}} + \cal A
\end{eqnarray}
where ${\cal{T}}$ and $\cal A$ are the marginal parts of
$\cal{T}^{\L}$ and $\cal A^{\L}$.

The additional renormalization conditions to which we refer in the following
sections,
have been chosen through a covariantization procedure of the
renormalization conditions of the corresponding vector theory with minimal
subtraction.
For example to the tensor $Tr[t^a_s S_i t^b_s S_j]$ we associate in
the chiral theory the tensor
$\frac{1}{4}Tr\,tr[t^a y^\dagger_i \bar{t}^b y_j]$, which has the same transformation
property.
According to this rule by construction $S_W$ does not contain monomials in the
gauge fields with the Levi-Civita tensor. This leads to the definition
of the anomaly in the left-right symmetric form.

The Bardeen anomaly can be obtained through different renormalization
condition, allowing in $S_W$ terms proportional to the Levi-Civita tensor.

\subsection{One loop}

At one loop using the tree level action (\ref{treeact}) and the rules
 (\ref{inswi1},\ref{inswi2}) for the insertions one computes:

\begin{equation}\label{tau}\bsp
{\cal{T}} &= \int \pa_\m \e^e \Big[
b_1^{ea} D_\n^{ab} \pa_\n A_\m^b +
b_4^{ec} f^{cab} A_\n^a (\pa_\m A_\n^b - \pa_\n A_\m^b) + \\ 
&\quad\quad(f^{ebx} b_6^{xy} f^{yac} + \frac{1}{3!} b^{eabc})
A_\m^a A_\n^b A_\n^c +  \\
&\quad\quad i \bar{\psi} \g_\m b^e \psi + 
b_{ij}^e \phi_i \pa_\m \phi_j +\oh b_{ij}^{ea} A_\m^a \phi_i \phi_j 
\Big]  - {\cal A} 
\esp\end{equation}
\begin{equation}\label{anoma}
{\cal A} = - \int \frac{2}{3} i \e^{\m\n\r\s} \pa_\m \e^e Tr \Big[ t_R^e (
\pa_\n A_{R \r} A_{R \s} + \oh A_{R \n} A_{R \r} A_{R \s}) - (R \to L )\Big] 
\end{equation}
where $A_{R \m} \equiv i t_R^a A_\m^a, A_{L \m} \equiv i t_L^a A_\m^a
$.

With the notations: 
\begin{equation}\bsp
&S_2(F)^{ab} = \oh Tr[t^a_R t^b_R+t^a_L t^b_L] \qquad;
\qquad S_2(S)^{ab} = \theta^a_{ij}\theta^b_{ji}\\
&S_4(F)^{abcd} = \oh Tr[t^{(a}_R t^b_R t^c_R t^{d)}_R+t^{(a}_L t^b_L t^c_L t^{d)}_L]\\
& S_4(S)^{abcd} = \theta^a_{ij}\theta^b_{jk}\theta^c_{kh}\theta^d_{hi}+
\theta^a_{ij}\theta^b_{jk}\theta^d_{kh}\theta^c_{hi}+
\theta^a_{ij}\theta^c_{jk}\theta^b_{kh}\theta^d_{hi}
\esp\end{equation}
the values of the coefficients in eq.(\ref{tau}) are:
\begin{equation}\bsp
&b_1^{ab} = - \frac{2}{3} S_2(F)^{ab} + \frac{1}{60} S_2(S)^{ab} \\
&b_4^{ab} = - \frac{5}{9} S_2(F)^{ab} +  \frac{1}{72} S_2(S)^{ab} \\
&b_6^{ab} = \frac{34}{105} S_2(F)^{ab} -\frac{7}{360} S_2(S)^{ab} \\
&b^{abcd} = 4 S_4(F)^{abcd} -\frac{11}{210} S_4(S)^{abcd} \\
&b^a = -\frac{53}{120} y_i^{\dagger} \bar{t}^a y_i +
\frac{7}{120}  y_i^{\dagger} y_j \theta^a_{ij}  \\
&b_{ij}^a = \frac{i}{4}Tr[Y_i Y_j^\dagger t^a_L +Y_i^\dagger Y_j
t^a_R] - (i \leftrightarrow j) \\
&b_{ij}^{ab} = \frac{121}{210}\left(Tr[\{t^a_R,t^b_R\}Y_i^\dagger Y_j
+\{t^a_L,t^b_L\}Y_i Y_j^\dagger] + (i \leftrightarrow j)\right) -  \\
&\quad\quad\:\:\frac{16}{105}\left(Tr[t^a_R Y_i^\dagger t^b_L Y_j + t^a_L Y_i t^b_R
Y_j^\dagger] + (i \leftrightarrow j) \right) 
\esp\end{equation}

To determine the Wilsonian renormalization conditions one must find 
$S_W$ ; choose them of the form 
\begin{equation}\label{sw}\bsp
S_W^{(A)} = \int &\oh a_1^{ab} A_\mu^a \pa^2 A_\mu^b +
\oh a_2^{ab} A_\mu^a \pa_\mu \pa_\nu A_\nu^b+
a_3^{ax} f^{xbc} \pa_\n A_\m^a A_\m^b A_\n^c +  \\
&\frac{1}{4} (f^{acx} a_4^{xy}  f^{ybd} + \frac{1}{6} a^{abcd})
 A_\mu^a A_\mu^b A_\nu^c A_\nu^d + 
i A_\m^a \bar{\psi} \g_\m a^a \psi +  \\
&A_\m^a \phi_i a_{ij}^a \pa_\m \phi_j +
\frac{1}{4} A_\m^a A_\m^b \phi_i \phi_j a_{ij}^{ab} 
\esp\end{equation}

The effective Ward identities give the following relations  between
the $S_W^{(A)}$ coefficients  and those for
${\cal{T}}$ : 
\begin{align}\label{alwi}
&(a_1 + a_2 + b_1)^{ab} = 0 & 
&(a_1 + a_3 - b_4)^{ab} = 0 \no \\
&(a_3 - a_4 - b_6)^{ab} = 0 &
&(a+b)^{abcd} = 0 \no \\
&t^e a - a^e + b^e = 0 & &[t^e,a] = 0  \\
&(a^a + b^a - i\theta^a a )_{ij} = 0 &
&i a_{ik}^a \theta_{kj}^b + i a_{jk}^a \theta_{ki}^b - a_{ij}^{ab}
- b_{ij}^{ba} = 0 \no
\end{align}

Using (\ref{alwi}), a set of independent parameters of $S_W^{(A)}$ is
$a,a_{ij}$ and $a_2^{ab}$. The former two parameters have been fixed
in the previous sections, to be compatible with minimal subtraction in
the non-chiral case; making the analogous choice for the new parameter
we get
\begin{eqnarray}
a_2^{ab} = \frac{8}{5} S_2(F)^{ab} + 
\frac{11}{60}  S_2(S)^{ab} + z^{ab}
\end{eqnarray}
with $z^{ab} = 0$ for the minimal subtraction choice.
We have checked all the relations (\ref{alwi}) by explicit computation
of the marginal Wilsonian Green functions and the corresponding 
${\cal{T}}$ insertions.

Having determined the renormalization conditions, we can now compute
the one-loop bare action:
\begin{equation}\bsp
&S_{(1)} =  \int \frac{1}{4} F_{\m\n}^a 
[ - \frac{8}{3 \e} S_2(F)^{ab} - \frac{1}{3 \e} S_2(S)^{ab} + z^{ab}]
 F_{\m\n}^b +\\
&\qquad\qquad S^{\f NDR}_{(1)}+S^{\p NDR}_{(1)}+\Delta S_{(1)}
\esp \end{equation}
where
$F_{\m \n}^a \equiv \pa_\m A_\n^a - \pa_\n A_\m^a - f^{abc} A_\m^b
A_\n^c$,~ $S^{\f NDR}_{(1)}$ and $S^{\p NDR}_{(1)}$ are the
covariantization
of the kinetic terms for fermions and scalars in absence of the gauge field;
$\Delta S_{(1)}$ is the non-naive part of $S_{(1)}$.

${\cal A}$ in (\ref{anoma}) is the Adler-Bardeen anomaly \cite{AB};
for a previous derivation using Wilsonian methods, see \cite{BV}. 

In order to compute the Bardeen anomaly in the next subsection we
write $\Delta S^{(A)}_{(1)}$, the gauge-dependent part of $\Delta S_{(1)}$,  with the variables ${\cal A}_\m$ and
${\cal V}_\m$ so defined:
\begin{equation}\label{VA}
A_\m = {\cal V}_\m + \gc {\cal A}_\m \quad;\quad
{\cal V}_\m = i t_s^a A_{\m}^a \quad;\quad
{\cal A}_\m = i t_p^a A_{\m}^a  
\end{equation}

As a consequence:
\begin{eqnarray}\label{FVA}
&&F_{\m\n} = \pa_\m A_\n - \pa_\n A_\m + [A_\m,A_\n] = 
{\cal V}_{\m\n} + \gc {\cal A}_{\m\n} \no \\
&&{\cal V}_{\m\n} = \pa_\m {\cal V}_\n -\pa_\n {\cal V}_\m +
[{\cal V}_\m,{\cal V}_\n] + [{\cal A}_\m,{\cal A}_\n]  \\
&&{\cal A}_{\m\n} = D^V_\m {\cal A}_\n -D^V_\n {\cal A}_\m \no
\end{eqnarray}
where we define $
D^V_\m f = \pa_\m f + [{\cal V}_\m,f] $

A similar decomposition for the scalars fields is introduced:
\begin{equation}\label{scsb}
\s \equiv S_i \phi_i \quad;\quad \pi\equiv P_i \phi_i  
\end{equation}

$\Delta S^{(A)}_{(1)}$ is then cast in the form:
\begin{equation}\label{dsb}\bsp
&\Delta S^{(A)}_{(1)} = 4 \int Tr \Big[ 
\frac{1}{6} (D_\m^V {\cal A}_\n)^2 + 
{\cal V}_{\m\n} {\cal A}_\m {\cal A}_\n -
\frac{1}{3} [{\cal A}_\m, {\cal A}_\n]^2 -
\frac{1}{3} {\cal A}_\m^2 {\cal A}_\n^2 + \\
&\quad\frac{1}{3}\e^{\m\n\r\s}\left( \pa_\m {\cal V}_\n {\cal V}_\r {\cal A}_\s +
{\cal V}_\m \pa_\n {\cal V}_\r {\cal A}_\s +
\frac{3}{2} {\cal V}_\m {\cal V}_\n {\cal V}_\r {\cal A}_\s +\oh  {\cal V}_\m {\cal A}_\n {\cal A}_\r {\cal A}_\s\right)
\ - \\&\quad  \frac{1}{6} (D_\m^V \pi)^2 +
i {\cal A}_\m \{ \pi, D_\m^V \s \} + {\cal A}_\m^2 (\s^2 + \pi^2)+
\oh \{ {\cal A}_\m ,\pi \}^2 \Big] + \\ &\quad
\int \bar{\psi} y_i \g_\m \gc {\cal A}_\m y_i \psi 
\esp\end{equation}

\subsection{Bardeen anomaly}
The theory we considered up to now is very general, so 
for instance one should be able to compute the Bardeen anomaly
\cite{Bardeen}.
This requires a suitable modification of the non
naive counterterm part  $\Delta S^{(A)}$.
Let's consider the case in which $G=\tilde{G}_R \times\tilde {G}_L$. In our model
this is made by choosing that half of the generators ${t^a_R}$ and half of
the generators ${t^a_L}$ vanish in such a way that $t^a_R t^b_L=t^a_L t^b_R=0$.
Let's consider moreover the case in which the left and the right
representations of $\tilde{G}$ are the same  so that  the vectorial
representation of $\tilde{G}$ turns out to be defined.

Decompose the gauge parameter as
\begin{eqnarray}
\e \equiv i t^a \e^a = \alpha + \gc \beta
\end{eqnarray}
Now ${\cal V}_\m$ and ${\cal A}_\m$ as well as $\alpha$ and $\beta$, which are
respectively the vectorial and axial gauge parameters, can be considered
as independent quantities.
The gauge transformation
\begin{eqnarray}
\d A_\m = - \pa_\m \e +[\e,A_\m]
\end{eqnarray}
decomposes into vectorial and axial gauge transformations.

Under  vectorial gauge transformations one has
\begin{eqnarray}
&&\d_{\alpha} {\cal V}_\m =  - D^V_\m \alpha
\quad;\quad \d_{\alpha} {\cal A}_\m =  [\alpha, {\cal A}_\m] 
\end{eqnarray}
and the scalar fields introduced in eq.(\ref{scsb}) transform
homogeneously.

Under  axial gauge transformations one has
\begin{eqnarray}
\d_{\beta} {\cal V}_\m = [\beta, {\cal A}_\m]\quad;\quad
\d_{\beta} {\cal A}_\m = - D^V_\m \beta 
\end{eqnarray}

Under a vectorial gauge transformation one gets 
\begin{eqnarray}\label{vgt}
\d_{\alpha} S_{bare} = {\cal A}(\alpha )
\end{eqnarray}
where ${\cal A}(\alpha )$ is the vectorial part of the anomaly (\ref{anoma})
which indeed can be decomposed as
\begin{eqnarray}\label{decan} 
{\cal A} = {\cal A}(\alpha ) + {\cal A}(\beta )
\end{eqnarray}

where
\begin{equation}
\bsp
{\cal A}(\xi) = -& \frac{2}{3} \int \e^{\m\n\r\s}
Tr \Big[\pa_\m \xi \Big(
\pa_\n A_{R \r} A_{R \s} + \oh A_{R \n}  A_{R \r} A_{R \s} -  \\
&\pa_\n A_{L \r} A_{L \s} - \oh A_{L \n}  A_{L \r} A_{L \s} \Big)\Big]
\esp
\end{equation}
On the other hand under an axial gauge transformation one gets 
\begin{equation}\label{agt}\bsp
&\d_{\beta} S^{(A)}_{(\e-part)} = 
{\cal A}(\beta) - {\cal A}_{Bardeen}  \\
&{\cal A}_{Bardeen} =
4 \int \e^{\m\n\r\s} Tr \Big\{\beta \Big[ 
\frac{1}{4} {\cal V}_{\m\n} {\cal V}_{\r\s} +
\frac{1}{12} {\cal A}_{\m\n} {\cal A}_{\r\s} -  \\
&\qquad\qquad\frac{2}{3} ({\cal A}_\m {\cal A}_\n {\cal V}_{\r\s} +
{\cal A}_\m {\cal V}_{\n\r} {\cal A}_\s +
{\cal V}_{\m\n}  {\cal A}_\r  {\cal A}_\s) + 
\frac{8}{3} {\cal A}_\m {\cal A}_\n  {\cal A}_\r  {\cal A}_\s 
\Big]\Big\}
\esp\end{equation}

One can then modify the renormalization of the product of the currents by
subtracting out all the terms depending on the Levi-Civita tensor in
the finite part of the bare action (\ref{dsb}) and correspondingly by
introducing their gauge variation in eq.(\ref{ewi}). Being ${\cal T}$
unchanged the effect of subtracting eq.(\ref{vgt}) is to recover the
vectorial gauge invariance, the subtraction of eq.(\ref{agt}) put
the anomaly in  the Bardeen form \cite{Bardeen}.

Observe that the $\e$-dependent part of (\ref{dsb}) is not specific of
the BMHV regularization, being fixed its gauge variations 
(\ref{vgt}, \ref{agt}). 

The $\e$-independent part of (\ref{dsb})
is specific of the  BMHV regularization; for instance it takes a
different expression in \cite{BMNT}, where a different renormalization
scheme is used. Using (\ref{dsb}) with 
${\cal V}_\m = \frac{i}{2} A_\m^a (t_R+t_L)$
and ${\cal A}_\m = \frac{i}{2} A_\m^a (t_R-t_L)$ 
and making the non-minimal  choice 
$ z^{ab} = \frac{5}{3} S_2 (F)^{ab}$ 
we find agreement with the finite counterterm computed in \cite{Martin}  
using the Bonneau identities \cite{Bonnid} .

\subsection{Two loop}
\begin{figure} 
\centering
	\subfigure[]{
		\label{fig:gauge:a}
        	\includegraphics[width=2cm]{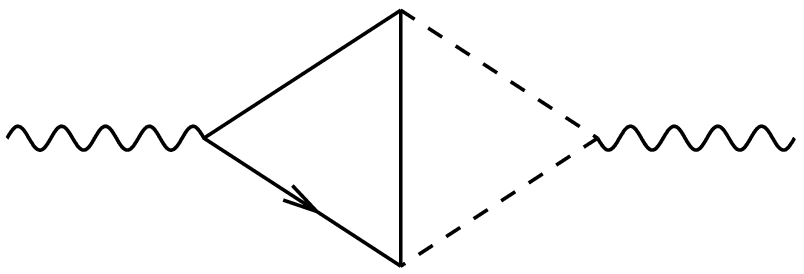}}
	\hspace{0.4cm}
	\subfigure[]{
 		\label{fig:gauge:b}
        	\includegraphics[width=2cm]{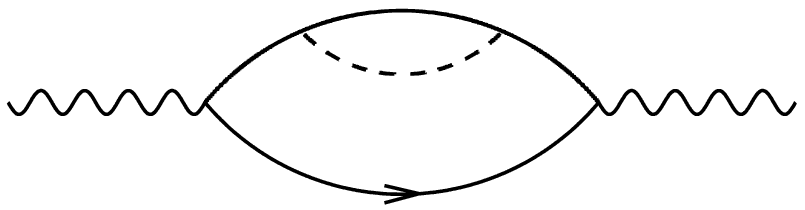}}
	\hspace{0.4cm}
	\subfigure[]{
		\label{fig:gauge:c}
		\includegraphics[width=2cm]{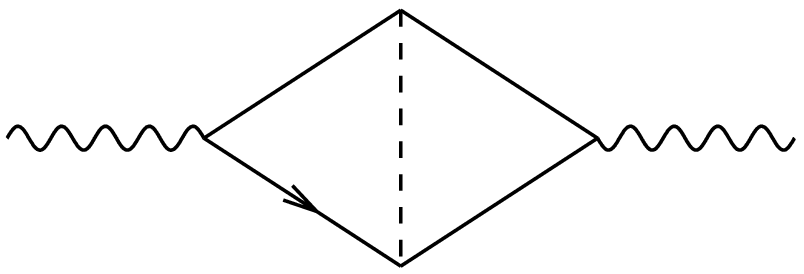}}
	\hspace{0.4cm}
	\subfigure[]{
		\label{fig:gauge:d}
		\includegraphics[width=2cm]{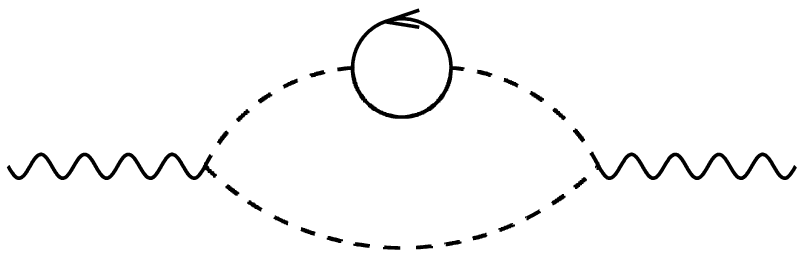}}
	\caption{Graphs of the two-point gauge field Green function.}
\label{fig:gauge}
\end{figure}

\begin{figure} 
\centering
	\subfigure[]{
		\label{fig:tgauge:a}
        	\includegraphics[width=2cm]{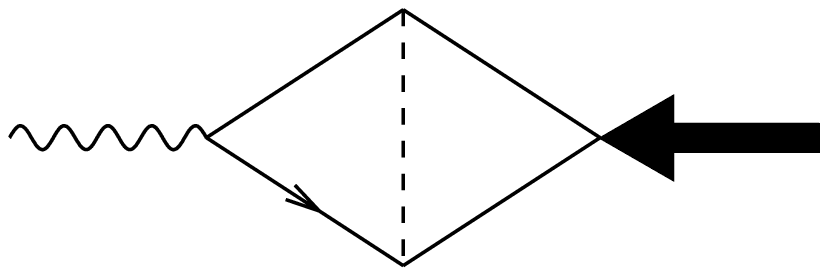}}
	\hspace{0.4cm}
	\subfigure[]{
 		\label{fig:tgauge:b}
        	\includegraphics[width=2cm]{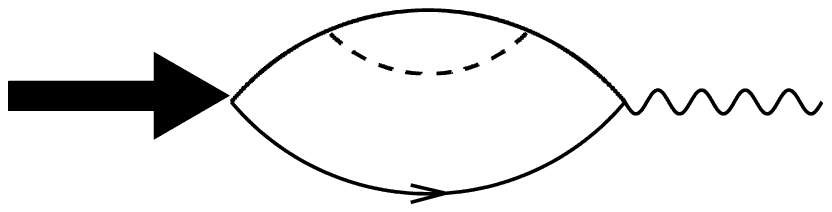}}
	\hspace{0.4cm}
	\subfigure[]{
		\label{fig:tgauge:c}
		\includegraphics[width=2cm]{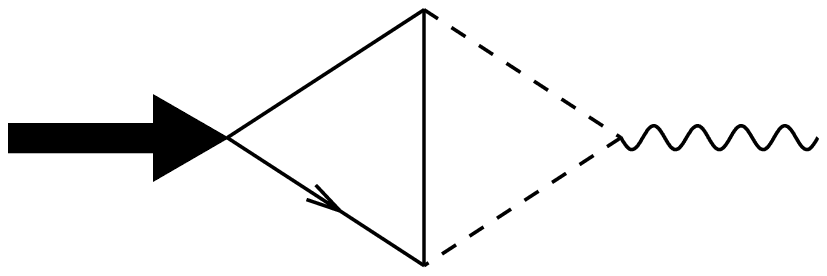}}
	\hspace{0.4cm}
	\subfigure[]{
		\label{fig:tgauge:d}
		\includegraphics[width=2cm]{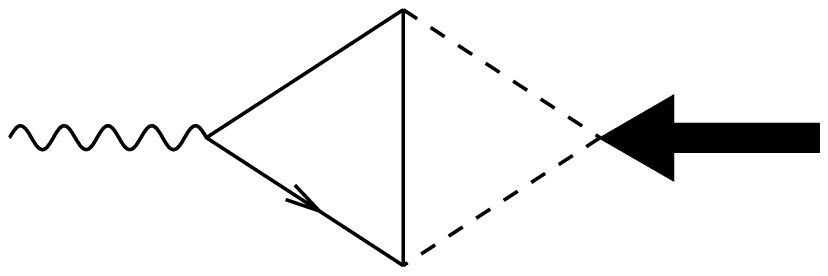}}
	\hspace{0.4cm}
	\subfigure[]{
		\label{fig:tgauge:e}
		\includegraphics[width=2cm]{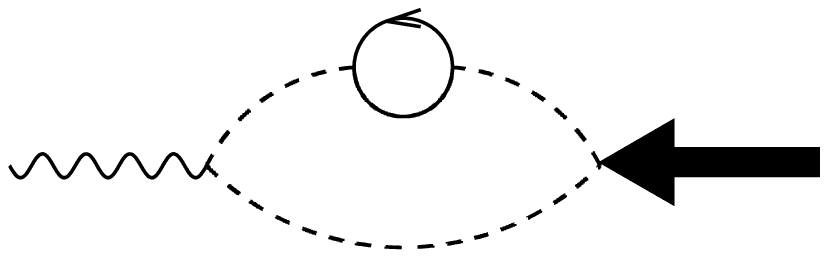}}
	\caption{Tau-gauge graphs}
\label{fig:tgauge}
\end{figure}

At two loops similar computations could be performed.
We have computed the marginal Wilsonian action in the two gauge field
sector, that is the coefficients $a_1^{ab}$ and $a_2^{ab}$ of
eq.(\ref{sw}) at two loop.
The graphs involved are shown in fig. \ref{fig:gauge}; their
contributions to $a_1^{ab}$ and $a_2^{ab}$ 
are expanded on the basis of invariant symmetric tensors:
\begin{equation}\label{basis}\bsp
&K_1^{ab} = \oh\left(tr\left[t^a_L Y_i t^b_R Y_i^\dagger\right]
 +tr\left[t^a_R Y_i^\dagger t^b_L Y_i\right]\right)\\
&K_2^{ab} = \oh\left(tr\left[t^a_L t^b_L Y_i Y_i^\dagger\right]
 +tr\left[t^a_R t^b_R  Y_i^\dagger Y_i\right]\right)
\esp\end{equation}
and are shown in table \ref{tab:gauge}.
In order to check the first Ward identity relation in (\ref{alwi}) we have also computed
the graphs shown in fig. \ref{fig:tgauge} and the results are
collected in table \ref{tab:tgauge}.
Actually we did not consider in figs. \ref{fig:gauge}, \ref{fig:tgauge}
the graph with only
scalar internal lines and quartic scalar vertex: it is finite and rigid
invariant, therefore it does not contribute to the bare action and it has to match separately its own ${\cal T}$ insertion.

\begin{table}
\begin{center}      
\begin{tabular}{|c|c|c|c||c|}
\hline
\multicolumn{2}{|c|}{\raisebox{2pt}{$a_1^{ab}$}}&\multicolumn{2}{|c||}{\raisebox{2pt}{$a_2^{ab}$}}&
\rule{0pt}{3ex}\raisebox{2pt}{graphs}\\
\hline
\rule{0pt}{3ex}\raisebox{2pt}{$K_1^{ab}$}&\raisebox{2pt}{$K_2^{ab}$}&
\raisebox{2pt}{$K_1^{ab}$}&\raisebox{2pt}{$K_2^{ab}$}&\\
\cline{1-4}
&&&&\\
$-\frac{593}{3645}+\frac{674}{2187}v$
& $0$
& $\frac{208}{729}-\frac{2066}{2187}v$
& $0$
& Fig.\ref{fig:gauge:a}\\ 
&&&&\\
$0$
&$\frac{57469}{51030}-\frac{98}{2187}v$ 
&$0$
&$\frac{38657}{25515}+\frac{98}{2187}v$
& Fig.\ref{fig:gauge:b}\\
&&&&\\
$\frac{13178}{3645}-\frac{9956}{2187}v$
& $-\frac{13178}{3645}+\frac{9956}{2187}v$
&$-\frac{9674}{3645}+\frac{3428}{2187}v$
& $\frac{9674}{3645}-\frac{3428}{2187}v$
&Fig.\ref{fig:gauge:c}\\ 
&&&&\\
$-\frac{80042}{25515} -\frac{1411}{2187}v$
& $\frac{80042}{25515} +\frac{1411}{2187}v$
&$\frac{195233}{51030}+\frac{13858}{2187}v$
& $-\frac{195233}{51030}-\frac{13858}{2187}v$
&Fig.\ref{fig:gauge:d}\\
&&&&\\
\hline
\end{tabular}
\end{center}
\caption{Contributions to the two-point gauge Green function: the numbers include the contributions of the two-loop graphs with the corresponding
one-loop subtraction and sum over external legs permutation}
\label{tab:gauge}
\end{table}

\begin{table}
\begin{center}      
\begin{tabular}{|c|c||c|}
\hline
\multicolumn{2}{|c||}{\raisebox{2pt}{$b_1^{ab}$}}&
\rule{0pt}{3ex}\raisebox{2pt}{graphs}\\
\hline
\rule{0pt}{3ex}\raisebox{2pt}{$K_1^{ab}$}&\raisebox{2pt}{$K_2^{ab}$}&\\
\cline{1-2}
&&\\
$-\frac{262}{1215}+\frac{748}{729}v$
& $0$
& Fig.\ref{fig:tgauge:a}\\ 
&&\\
$0$
& $\frac{1171}{2430}-\frac{248}{729}v$
& Fig.\ref{fig:tgauge:b}\\
&&\\
$\frac{253}{243}-\frac{1508}{729}v$
& $-\frac{253}{241}+\frac{1508}{729}v$
& Fig.\ref{fig:tgauge:c}\\
&&\\
$-\frac{157}{405}+\frac{268}{243}v$
& $\frac{157}{135}-\frac{268}{81}v$
& Fig.\ref{fig:tgauge:d}\\
&&\\
$\frac{1751}{1890}-\frac{124}{81}v$
& $-\frac{1751}{1890}+\frac{124}{81}v$
& Fig.\ref{fig:tgauge:e}\\
&&\\
\hline
\end{tabular}
\end{center}
\caption{Contributions to the one-point ${\cal T}$ Green function.}
\label{tab:tgauge}
\end{table}

From the pole part of graphs in fig. \ref{fig:gauge} we computed
  the gauge invariant part of $S^{(A)}$
  at two loops:

\begin{equation}\label{SA2}
S^{(A)}_{(2)} = \int \frac{1}{4} F_{\m\n}^a 
[ - \frac{2}{\e} K_2^{ab}]
 F_{\m\n}^b 
\end{equation}
finding agreement with \cite{Mac}.

From the non-invariant part we derived the following contribution
to the bare action:
\begin{equation}\label{dSA2}\bsp
&\Delta S^{(A)}_{(2)} =  \oh \int Tr\Big[
\left(\frac{2}{3\e}-\frac{53}{36}\right)
	\{{\cal A}_\m,P_i\}(D_\n^V)^2\{{\cal A}_\m,P_i\}+
	2{\cal A}_\m(D_\n^V)^2{\cal A}_\m P_iP_i -\\
&\qquad\qquad\left(\frac{2}{3\e}+\frac{13}{36}\right)
	\{{\cal A}_\m,S_i\}(D_\n^V)^2\{{\cal A}_\m,S_i\}+
	\frac{5}{3}{\cal A}_\m(D_\n^V)^2{\cal A}_\m S_iS_i\Big]+\\
&\quad\quad			\frac{1}{4}\int Tr\Big[
\left(\frac{8}{3\e}-\frac{13}{9}\right)({\cal A}_{\m\n}S_i)^2
-\frac{5}{9}({\cal A}_{\m\n})^2(S_i)^2 +
\left(\frac{4}{3\e}-\frac{22}{9}\right)({\cal A}_{\m\n}P_i)^2+\\
&\qquad\qquad\left(\frac{4}{3\e}-\frac{4}{9}\right)({\cal A}_{\m\n})^2(P_i)^2 
-\left(\frac{4}{9\e}+\frac{31}{27}\right)({\cal V}_{\m\n}P_i)^2 +\\
&\qquad\qquad\left(\frac{28}{9\e}-\frac{29}{27}\right)({\cal V}_{\m\n})^2(P_i)^2 +
i\left(\frac{8}{3\e}-\frac{13}{9}\right){\cal V}_{\m\n}S_i{\cal A}_{\m\n}P_i+\\
&\qquad\qquad i\left(-\frac{16}{9\e}+\frac{35}{27}\right){\cal V}_{\m\n}P_i{\cal A}_{\m\n}S_i+
i\left(\frac{4}{3\e}-\frac{1}{9}\right){\cal V}_{\m\n}{\cal
A}_{\m\n}P_iS_i-\\
&\qquad\qquad\frac{5}{9}i{\cal V}_{\m\n}{\cal A}_{\m\n}S_iP_i+
i\left(-\frac{4}{9\e}+\frac{14}{27}\right){\cal V}_{\m\n}P_iS_i{\cal A}_{\m\n}\Big]
\esp\end{equation}

We used the notation of eqs.(\ref{VA},\ref{FVA});
the r.h.s. of eq.(\ref{dSA2}) is not simply quadratic in the gauge
fields
because we completed it in order to have an expression gauge-invariant
under
vectorial transformation. This fact has been possible because the derivatives
$\pa_\m {\cal V}_\n $ of the gauge fields ${\cal V}_\m$ appear only in the
combination  $\pa_{[\m}{\cal V}_{\n]}$ as expected, since dimensional
regularization and ours renormalization condition respect vectorial symmetry.
 
 We checked the non renormalization theorem of the anomaly only in the
two-gluon sector of
eq. (\ref{ewi}).
We have computed the part of  ${\cal T}$ from which the 
anomaly might arise:
\begin{equation}
{\cal \bar T} = \int \pa_\m \e^a(x) \pa_\n A^b_\r A^c_\s \e^{\m\n\r\s} T_{abc}
\end{equation}
where $T_{abc}$ is an invariant tensor symmetric in the last two indices.  The non renormalization
 theorem
requires that the completely symmetric part
$T_{(abc)}$, which cannot be eliminated by a local counterterm, must
 vanish.

\begin{figure} 
\centering
\subfigure[]{\label{fig:an:a}\includegraphics[width=1.4cm]{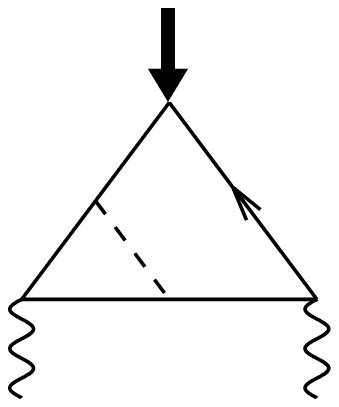}}
	\hspace{0.4cm}
\subfigure[]{\label{fig:an:b}\includegraphics[width=1.4cm]{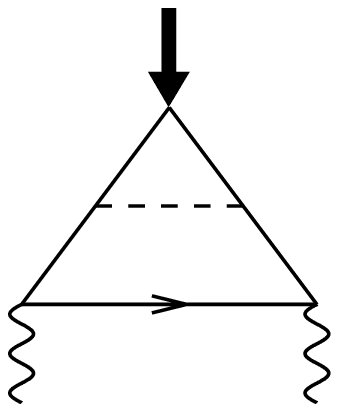}}
	\hspace{0.4cm}
\subfigure[]{\label{fig:an:c}\includegraphics[width=1.4cm]{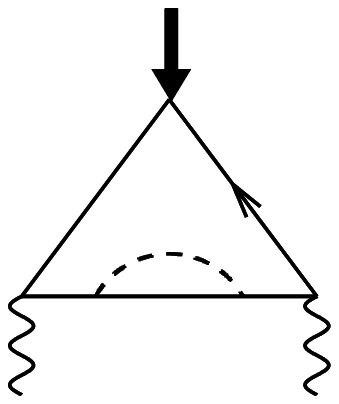}}
	\hspace{0.4cm}
\subfigure[]{\label{fig:an:d}\includegraphics[width=1.4cm]{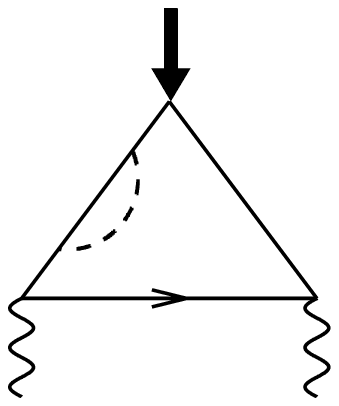}}
	\hspace{0.4cm}
\subfigure[]{\label{fig:an:e}\includegraphics[width=2.1cm]{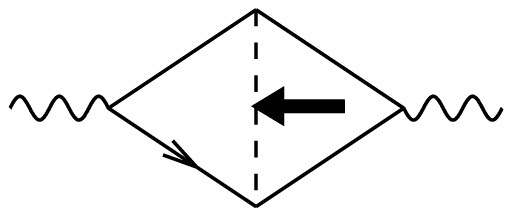}}
	\hspace{0.4cm}
\subfigure[]{\label{fig:an:f}\includegraphics[width=1.4cm]{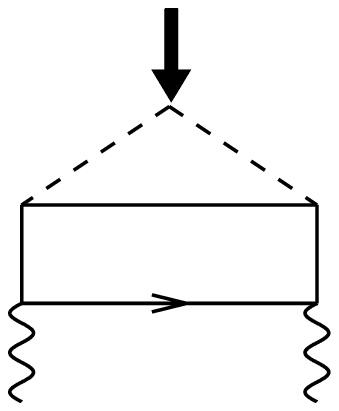}}
	\hspace{0.4cm}
\subfigure[]{\label{fig:an:g}\includegraphics[width=2.1cm]{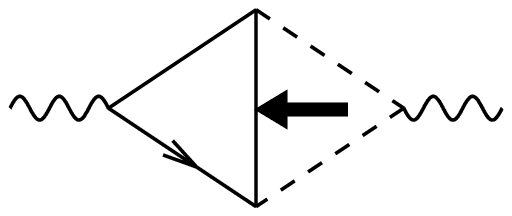}}
	\hspace{0.4cm}
\subfigure[]{\label{fig:an:h}\includegraphics[width=2.1cm]{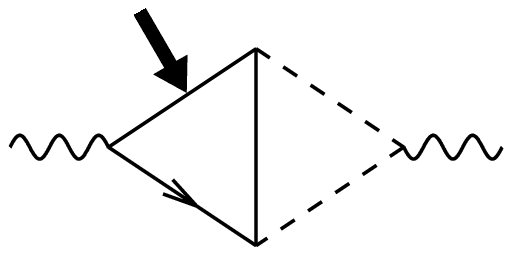}}
\caption{Two-loop graphs contributing to the anomaly}
\label{fig:anom}
\end{figure}

\begin{table}
\centering 
\begin{tabular}{|c|c||c|}
\hline
\rule{0pt}{3ex}          
\raisebox{2pt}{$T_{abc}^{(1)} $} &\raisebox{2pt}{$T_{abc}^{(2)} $} &\raisebox{2pt}{ graphs}\\
 \hline 
&&\\         
$-\frac{6797}{21870}i-\frac{4688}{6561}iv$
& $ 0 $ & Fig. \ref{fig:an:a}\\ 
&&\\
$\frac{18119}{21870}i-\frac{12880}{6561}iv$
& $ 0 $ & Fig. \ref{fig:an:b}\\ 
&&\\ 
$0$ & $-\frac{55441}{204120}i-\frac{76}{2187}iv $ & Fig. \ref{fig:an:c}\\ 
&&\\
$0$ & $-\frac{10931}{204120}i-\frac{308}{2187}iv $ & Fig. \ref{fig:an:d}\\ 
&&\\ 
$-\frac{254}{243}i+\frac{2080}{729}iv$ & $ 0 $ & Fig. \ref{fig:an:e}\\
&&\\
$\frac{98}{135}i-\frac{440}{243}iv $
& $ -\frac{98}{135}i+\frac{440}{243}iv $ & Fig. \ref{fig:an:f}\\ 
&&\\
$\frac{4}{3}i $ 
& $0 $ & Fig. \ref{fig:an:g}\\
&&\\
$-\frac{1861}{1215}i+\frac{1192}{729}iv $
& $\frac{1861}{1215}i-\frac{1192}{729}iv  $ & Fig. \ref{fig:an:h}\\ 
&&\\
\hline
\end{tabular}
\caption{Two-loop coefficients for the anomaly}
\label{tab:anom}
\end{table}

Fig. \ref{fig:anom} shows the graphs which give
contributions to $T_{(abc)}$. Each contribution is decomposed on a
suitable basis of symmetric tensor, we have chosen a basis in which 
$\theta^a_{ij}$ never appears explicitly (symmetrization in the
indices $a,b,c$ is understood):
\begin{equation}\bsp
& T_{abc}^{(1)} = Tr[Y^\dagger_k Y_k t^a_R t^b_R t^c_R] - Tr[Y_k Y^\dagger_k t^a_L t^b_L t^c_L ]\\
& T_{abc}^{(2)} = Tr[Y^\dagger_k t^a_L Y_k t^b_R t^c_R]-Tr[ Y_k t^a_R Y^\dagger_k t^b_L t^c_L]
\esp\end{equation}

The results of our calculations are summarized in table
\ref{tab:anom}: as expected the sum of the contributions of every column
vanishes, so that the Adler-Bardeen theorem is verified.

One could address the question whether $T_{abc}$ and not only its symmetric
part is actually vanishing. For $G=SU(N)$, using the Young tableaux
one can easily prove that invariant tensors $X_{abc}$
with mixed symmetry do not exist; \footnote{We thank C. Destri for explaining this
point to us.} for 
more general groups see \cite{Okubo}.

\section{Concluding remarks}

We have shown that Wilsonian methods are useful in the BMHV dimensional
renormalization of  theories with chiral symmetries;  the two-loop
renormalization of the most general Yukawa theory becomes straightforward in this
scheme, while in a more conventional approach, based on the
verification
of usual Ward identities, it is a non trivial task.

In our approach  
one renormalizes  the effective Wilsonian action along a flow on which the
subtraction integrals are easily computed using standard techniques.

There is some arbitrariness in the choice of this flow; we worked with
the $n=2$ flow, which is the
simplest one from a computational viewpoint: it is quite obvious that
calculations with the $n \geq 3$ flows are more complicated, although the
bare action is the same. On the other hand the $n=1$ 
flow, which coincides with the
auxiliary mass method, does not respect renormalizability
so the procedure described in this paper fails down.
It is a convenient method
in  nonchiral theories where minimal subtraction respects the Ward
identities.

In the future we intend to apply  our formalism to chiral gauge theories.

\newpage

\section*{Appendix. Two-loop bare Yukawa action}

We have written the naive part of the counterterms in the Tables;
here we report the remaining counterterms,
necessary to recover the rigid chiral
symmetry in the BMHV scheme.

The two-loop scalar quadratic counterterm is

\begin{equation*}\label{bb}\bsp
&\Delta c_{ij}(p) =  c_{ij}^{\phi} p^2 + c_{ij}^{m^2, kl}(n) m^2_{kl}
+ \hat{p}^2 terms\\
&c_{ij}^{\phi} = Tr \Big[ 2 S_i P_k S_j P_k + 
\left(-\frac{8}{9}+\frac{20}{3 \e}\right) P_i S_k P_j S_k+ 
\left(\frac{23}{9}-\frac{20}{3 \e}\right) (S_i S_k P_j P_k + \\
&\quad\quad S_i P_k P_j S_k + P_i P_k P_j P_k) +
\left(1-\frac{4}{\e}\right) S_i S_j P_k P_k +  \\
&\quad\quad\left(\frac{7}{9}-\frac{4}{3 \e}\right)(S_i S_k P_k P_j +S_i P_j P_k S_k+
P_i P_j S_j S_k) +\left(\frac{13}{9}-\frac{16}{3 \e}\right) P_i P_j P_j P_k
\Big]\\
&c_{ij}^{m^2, kl} = Tr[  
(3-\frac{4}{\e})(S_j S_k P_i P_l+ S_j P_k P_i S_l + P_j S_k S_i
P_l+ P_j P_k S_i S_l) -  \\
&\quad\quad 8 P_j S_k P_i S_l +(\frac{64}{9}-\frac{32}{3 \e}) P_j P_k P_i P_l]
+ (\frac{80}{9}-\frac{64}{3 \e}) P_j P_k P_l P_i +
 \\
&\quad\quad(2-\frac{8}{\e})(S_j S_k P_l P_i+ S_j P_k P_l S_i + P_j S_k
S_l P_i + P_j P_k S_l S_i) + 
(\frac{2}{\e}-\frac{7}{6}) P_m P_l ] h_{ijmk} 
\esp\end{equation*}
and the symmetrization in $i,j$ is understood.

The two-loop fermionic quadratic counterterm is
\begin{equation*}\label{ff}\bsp
&\Delta c(p) = \ds{p}\, c_{\psi} + [\hat{\ds{p}}-terms] \\
&c_{\psi} = i y_j^{\dagger}[
(\frac{1}{36} - \frac{1}{3 \e}) P_i P_i  y_j +
 (\frac{1}{8}-\frac{1}{2 \e}) i \gc
(P_i y_j^{\dagger} - y_i P_j) y_i +
(\frac{43}{54}-\frac{10}{9 \e}) Tr(P_i P_j) y_i] 
\esp\end{equation*}

The two-loop fermion-fermion-scalar counterterm is
\begin{equation*}\label{fb}\bsp
&\Delta c_i = \sum_{n=1}^7 c_i(n)  \\
&c_i(1) = i y_k [  
(-\frac{7}{6}+\frac{2}{\e}) Tr(P_j P_k) S_i +  \\
&\quad\quad i \gc (\frac{1}{2}-\frac{2}{\e})Tr(S_j S_k) P_i+
i \gc (\frac{22}{27}-\frac{32}{9 \e})Tr(P_j P_k) P_i] y_j \rule{2cm}{0pt}\\
&c_i(2) = i y_k [ 
i \gc (\frac{1}{4}-\frac{1}{\e})(S_j S_i P_j + P_j S_i
S_j)+\\
&\quad\quad i \gc \frac{2}{\e} S_j P_i S_j +
(-\frac{3}{4}+\frac{1}{\e}) (S_j P_i P_j +P_j P_i S_j)+
i \gc(\frac{2}{3}-\frac{2}{\e}) P_j P_i P_j ] y_k\\
&c_i(3) = i y_k [
i \gc(\frac{1}{8}-\frac{1}{2 \e})
( S_i Y_j P_j +P_i Y_j S_j+ \\
&\quad\quad S_j Y_j P_i+P_j Y_j S_i) +
i \gc(\frac{1}{9}-\frac{4}{3 \e})(P_i P_j P_j + P_j P_j
P_i)] y_k 
\esp\end{equation*}
\begin{equation*}\bsp
&c_i(4) = i y_k[ 
i \gc(\frac{1}{4}-\frac{1}{\e}) (S_i Y_j P_k +P_j Y_k S_i)-
i \gc(\frac{1}{4}+
\frac{1}{\e}) (P_i Y_j S_k +S_j Y_k P_i)- \\
&\quad\quad \frac{2}{\e} (P_i S_j P_k +P_j S_k P_i)-
i \gc(\frac{2}{3}+\frac{2}{\e})(P_i P_j P_k +P_j P_k P_i)]
y_j \\
&c_i(5) = i y_k [i \gc (2 S_j P_i S_k -\frac{1}{2} S_j y_i P_k-
\frac{2}{3} P_j P_i P_k -\frac{1}{2} P_j y_i S_k)] y_j\\
&c_i(6) = 0  \\
&c_i(7) = - \frac{1}{2} y_j P_k y_l h_{ijkl}
\esp\end{equation*}
The two-loop quartic scalar counterterm is
\begin{equation*}\label{b4}\bsp
&\Delta c_{ijkl} = \sum_{n=1}^9 c^{(2)}_{ijkl}(n)  
\\
&c(1) = 16~ Tr Y_m [
i (\frac{8}{\e}-\frac{2}{3}) S Y_m P^3 + 
i (\frac{3}{\e}-\frac{3}{4}) S Y_m (S Y P + P Y S)+
\\
&\quad\quad(\frac{8}{\e}+\frac{13}{3}) P Y_m P^3 +
P Y_m^{\dagger} (S Y^{\dagger}P+ P^3+PYS)]+
\\
&\quad\quad(\frac{6}{\e}-1) P Y_m (S Y P + P Y S) +
i (\frac{3}{\e}-\frac{9}{4})P Y_m ( S Y S - P Y P)+p.c.
\\
&c(2) = 4~ Tr Y_m [
\frac{3}{2}(Y^{\dagger} Y Y_m^{\dagger} Y Y^{\dagger}-
Y^2 Y_m Y^2)-
\\
&\quad\quad 8 P^2 Y_m (Y^2 + \frac{11}{8} P^2)+
(8 Y^{\dagger} Y - P^2) Y_m^{\dagger} P^2] + p.c. \\
&c(3) = Tr Y_m [
(\frac{12}{\e}- 3) (Y^4 Y_m - (Y^{\dagger} Y)^2 Y_m^{\dagger}) 
+ 12 P^4 Y_m -(\frac{64}{\e}+\frac{20}{3}) P^4 Y_m^{\dagger} + 
\\
&\quad\quad(\frac{32}{\e}-\frac{8}{3}) (2 i S Y  P^2 P_m +2 i P S^2 P P_m
- P^2 Y^{\dagger} S Y_m^{\dagger} +P^2 Y S Y_m) ]  + p.c. \\
&c(4) = Tr [ 
-(\frac{48}{\e}-12) (S_m S P P_n + S_m P^2 S_n +
\\
&\quad\quad P_m S^2 P_n + P_m P S S_n) -
(\frac{128}{\e}-\frac{160}{3}) P_m P^2 P_n ] h_{mn} 
 \\
&c(5) = Tr[ 
- (\frac{64}{\e}-\frac{128}{3})P_m P P_n P -48 S_m P S_n P- 
\\ 
&\quad\quad(\frac{24}{\e}-18)
(S_m S P_n P+S_m P P_n S+P_m S S_n P+P_m P S_n S)]h_{mn}  
\\
&c(6) = Tr[
(\frac{6}{\e}-\frac{7}{2}) P_m P_n] h_{mp} h_{np}
\\
&c(7) = c(8) = c(9) = 0 
\esp\end{equation*}
where the indices $i,j,k,l$, which are totally symmetrized, are
understood. \\$p.c.$ represents the terms obtained with the
substitution rule:  
$P \to -P$, $Y \to Y^\dagger$, $Y^\dagger \to Y $.

\end{document}